\documentclass[sigconf]{acmart}

\AtBeginDocument{%
  \providecommand\BibTeX{{%
    \normalfont B\kern-0.5em{\scshape i\kern-0.25em b}\kern-0.8em\TeX}}}

\copyrightyear{2023}
\acmYear{2023}
\setcopyright{rightsretained}
\acmConference[KDD '23]{Proceedings of the 29th ACM SIGKDD Conference on Knowledge Discovery and Data Mining}{August 6--10, 2023}{Long Beach, CA, USA}
\acmBooktitle{Proceedings of the 29th ACM SIGKDD Conference on Knowledge Discovery and Data Mining (KDD '23), August 6--10, 2023, Long Beach, CA, USA}
\acmDOI{10.1145/3580305.3599930}
\acmISBN{979-8-4007-0103-0/23/08}
\usepackage{makecell}
\usepackage{diagbox}
\usepackage{graphicx}
\usepackage{verbatim}
\usepackage{multirow}
\usepackage{algorithmic}
\usepackage{hyperref}
\usepackage{longtable}
\usepackage{array}
\usepackage{subfigure}
\usepackage{stfloats}
\usepackage{multicol}
\usepackage{color}
\usepackage{epstopdf}
\usepackage{bm}
\usepackage{amsmath}
\usepackage{booktabs} 
\usepackage{epsfig}
\usepackage{enumitem}
\usepackage{cleveref}
\usepackage{arydshln}
\usepackage{balance}
\usepackage{xspace}
\usepackage{enumitem}
\usepackage{listings}
\usepackage{xcolor}
\definecolor{codegreen}{rgb}{0,0.6,0}
\definecolor{codegray}{rgb}{0.5,0.5,0.5}
\definecolor{codepurple}{rgb}{0.58,0,0.82}
\definecolor{backcolour}{rgb}{1,1,1}

\lstdefinestyle{mystyle}{
    backgroundcolor=\color{backcolour},   
    commentstyle=\color{codegreen},
    keywordstyle=\color{magenta},
    numberstyle=\tiny\color{codegray},
    stringstyle=\color{codepurple},
    basicstyle=\ttfamily\footnotesize,
    breakatwhitespace=false,         
    breaklines=true,                 
    captionpos=b,                    
    keepspaces=true,                 
    numbers=left,                    
    numbersep=5pt,                  
    showspaces=false,                
    showstringspaces=false,
    showtabs=false,                  
    tabsize=2
}

\newcommand{\tabincell}[2]{\begin{tabular}{@{}#1@{}}#2\end{tabular}}

\newcommand{\hide}[1]{} 
\newcommand{\vpara}[1]{\vspace{0.05in}\noindent \textbf{#1 }}
\newcommand{\ipara}[1]{\vspace{0.03in}\noindent \textit{#1 }}

\newcommand{\secref}[1]{Section~\ref{#1}} 
\newcommand{\figref}[1]{Figure~\ref{#1}} 
\newcommand{\beq}[1]{\vspace{-0.03in}\begin{equation}#1\end{equation}\vspace{-0.03in}}

\newtheorem{problem}{Problem}
\newtheorem{definition}{Definition}

\newcommand{\da}{WhoIsWho}
\newcommand{\sda}{WhoIsWho\space}

\begin{document}


\title{Web-Scale Academic Name Disambiguation: \\
the \sda Benchmark, Leaderboard, and Toolkit}

\author{Bo Chen}
\affiliation{%
  \institution{Tsinghua University}
  \country{}
  }
\email{cb21@mails.tsinghua.edu.cn}

\author{Jing Zhang}
\authornote{Jing Zhang and Jie Tang are the corresponding authors.}
\affiliation{%
  \institution{Renmin University of China}
  \country{}
  }
\email{zhang-jing@ruc.edu.cn}

\author{Fanjin Zhang}
\affiliation{%
  \institution{Tsinghua University}
  \country{}
  }
\email{zfj17@mails.tsinghua.edu.cn}

\author{Tianyi Han}
\affiliation{%
  \institution{Zhipu.AI}
  \country{}
  }
\email{tianyi.han@aminer.cn}

\author{Yuqing Cheng}
\affiliation{%
  \institution{Zhipu.AI}
  \country{}
  }
\email{yuqing.cheng@aminer.cn}

\author{Xiaoyan Li}
\affiliation{%
  \institution{Zhipu.AI}
  \country{}
  }
\email{xinyan.li@aminer.cn}

\author{Yuxiao Dong}
\affiliation{%
  \institution{Tsinghua University}
  \country{}
  }
\email{yuxiaod@tsinghua.edu.cn}

\author{Jie Tang}
\affiliation{%
  \institution{Tsinghua University}
  \country{}
  }
\email{jietang@tsinghua.edu.cn}
\authornotemark[1]

\renewcommand{\shortauthors}{Chen, et al.}






\received{20 February 2007}
\received[revised]{12 March 2009}
\received[accepted]{5 June 2009}

\begin{abstract}

Name disambiguation---a fundamental problem in online academic systems--is now facing greater challenges with the increasing growth of research papers. For example, on AMiner, an online academic search platform, about 10\% of names own more than 100 authors. 
Such real-world challenging cases have not been effectively addressed by existing researches due to the small-scale or low-quality datasets that they have used.
The development of effective algorithms is further hampered by a variety of tasks and evaluation protocols designed on top of diverse datasets. To this end, we present \sda owning, a large-scale benchmark with over 1,000,000 papers built using an interactive annotation process, a regular leaderboard with comprehensive tasks, and an easy-to-use toolkit encapsulating the entire pipeline as well as the most powerful features and baseline models for tackling the tasks.
Our developed strong baseline has already been deployed online in the AMiner system to enable daily arXiv paper assignments\footnote{The public leaderboard is available at \href{http://whoiswho.biendata.xyz/}{http://whoiswho.biendata.xyz/}. The toolkit is at \href{https://github.com/THUDM/WhoIsWho}{https://github.com/THUDM/WhoIsWho}. The online demo of  daily arXiv paper assignments is at \href{https://na-demo.aminer.cn/arxivpaper}{https://na-demo.aminer.cn/arxivpaper}.}.

\end{abstract}
\maketitle

\begin{figure}[t]
	\centering
	\includegraphics[width=0.48\textwidth]{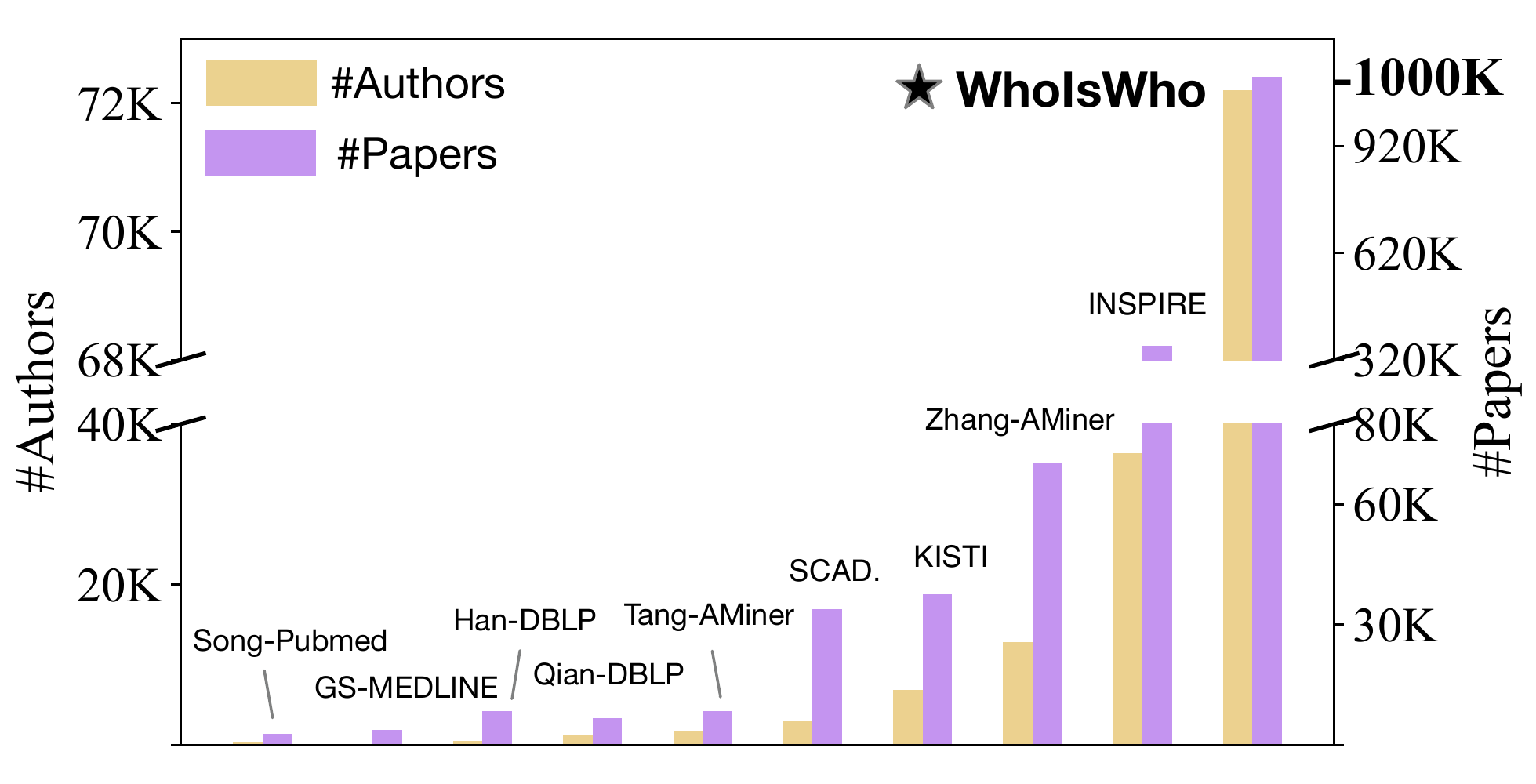}
	\vspace{-20pt}
	\caption{\label{fig:data} \textbf{The sizes of the prevailing name disambiguation benchmarks. }\textmd{Among these, \sda is the largest one with 1,000+ names, 70,000+ authors, and 1,000,000+ papers.} }
	\vspace{-13pt}
\end{figure}
\section{Introduction}
\label{sec:intro}
\begin{figure}[t]
	\centering
	\includegraphics[width=0.48\textwidth]{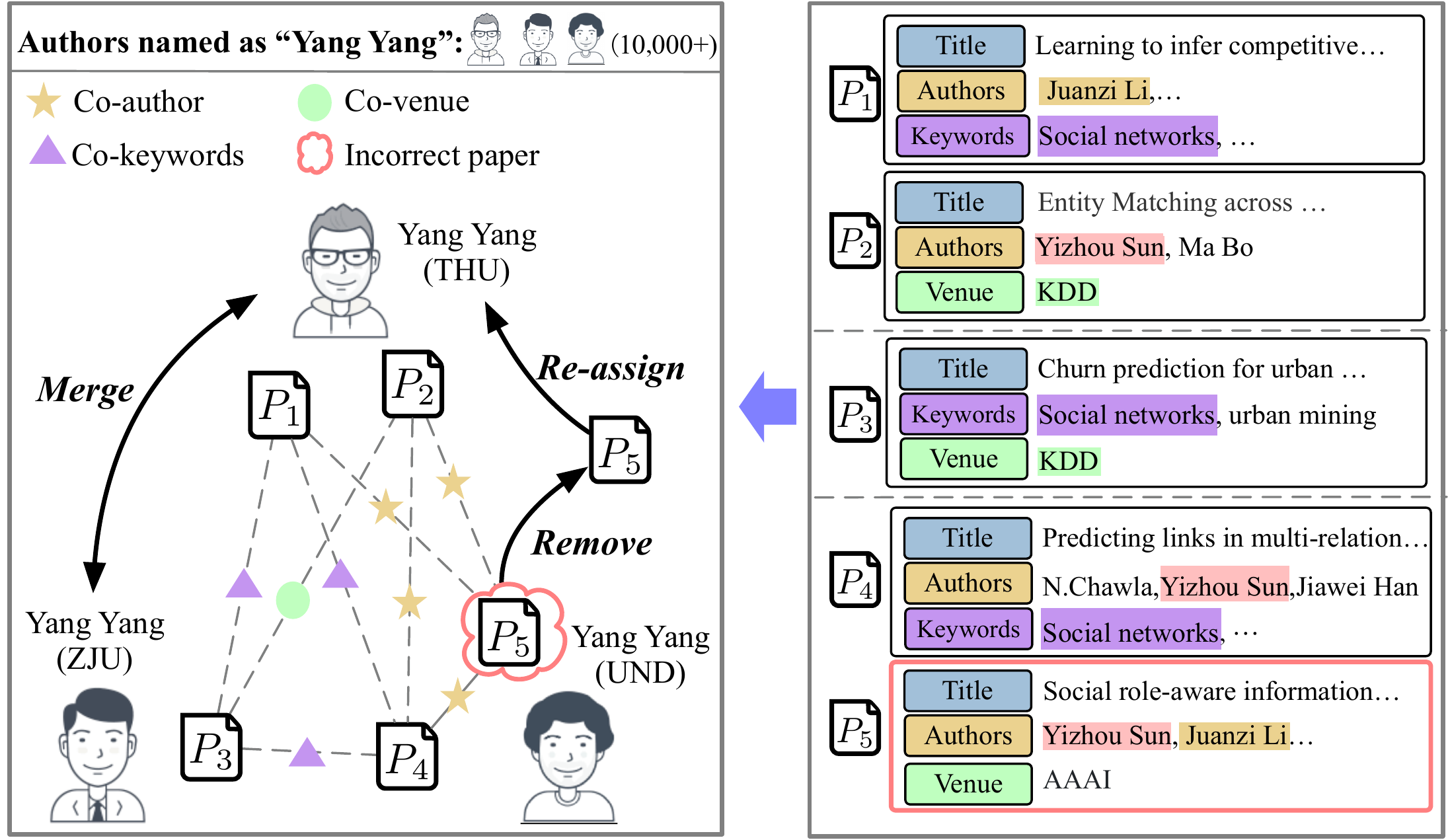}
	\vspace{-20pt}
	\caption{\label{fig:whoiswho_demo} Illustration of the challenges for annotating authors with the name ``Yang Yang''. \textmd{Paper $p_5$ is incorrectly assigned because of the coauthorship with the same third person. Two authors are mistakenly separated due to the organization shift. 
	}}
	\vspace{-5pt}
\end{figure}

Name disambiguation, aiming to clarify who is who, is one of the fundamental problems in online academic systems such as Google Scholar\footnote{https://scholar.google.com/}, 
Semantic Scholar\footnote{https://www.semanticscholar.org/},
and AMiner\footnote{https://aminer.org}. The past decades have witnessed a huge proliferation of research papers in all fields of science. For example, Google Scholar, Bing Academic Search, and AMiner have all indexed about 300 million papers~\cite{gusenbauer2019google,wang2020microsoft,tang2008arnetminer}. 
As a result, the author name ambiguity problem---the same authors with different name variants, or the different authors with the exact same name or homonyms---has become increasingly sophisticated in modern digital libraries.
For example, as of January 2023, there were over 10,000 authors with the name ``Yang Yang'' on AMiner. Three of them are displayed in \figref{fig:whoiswho_demo}. 
Since all three authors are computer scientists, there are intricate connections between their papers. 
Paper $P_5$, which belongs to ``Yang Yang(THU)'', was mistakenly assigned to ``Yang Yang(UND)'', because both ``Yang Yang'' coauthored with ``Yizhou Sun'', leading to the appearance of reliable co-author and co-keyword relationships between $P_5$ and the correct paper $P_4$ of ``Yang Yang(UND)''.
Furthermore, ``Yang Yang(THU)'' and ``Yang Yang(ZJU)'' are the same person but are separated into  two different authors due to organization shifts after graduation.
This real-world example demonstrates the great challenges of name disambiguation in online academic systems, which, however, can not be addressed by existing efforts ~\cite{louppe2016ethnicity,tang2012unified,wang2011adana,zhang2018name, chen2020conna, wang2020author, luo2020collective, li2021disambiguating, zhang2020strong, zheng2021dual}, because of the small-scaled low-quality benchmark and non-uniform task designs with evaluation settings.
In particular, even though several name disambiguation benchmarks,
such as PubMed~\cite{zhang2020mining, zeng2020large}, MAG~\cite{zhang2021lagos}, DBLP~\cite{kim2018evaluating}, etc.~\cite{wang2018acekg, kim2021orcid}, have been directly harvested from existing digital libraries, inevitably spurious information and assignment mistakes, as shown in \figref{fig:whoiswho_demo}, are detrimental to build effective algorithms~\cite{chen2022gccad, zhang2021name}.
In light of this, others attempt to manually annotate a small amount of high-quality data from the online noisy data in order to reduce the negative impact of these noises~\cite{song2015exploring, vishnyakova2019new, han2005name}.
However, as illustrated in \figref{fig:data}, the majority of them lack an adequate number of instances. 
Additionally, on top of these benchmarks,  
previous efforts have defined a variety of tasks and evaluation protocols, preventing us from fairly comparing different methods to promote the development of the name disambiguation community.

\vpara{Present Work.}
We present \da, a benchmark, a leaderboard, together with a toolkit for web-scale academic name disambiguation. Specifically, \sda has the following characteristics:

\noindent $\bullet$ \textbf{Interactive large-scale benchmark construction.} 
To create a challenging benchmark, we devise an interactive annotation process to label paper-author affiliations under a single name with high ambiguity with the aid of the developed visualization tool. 10+ professional annotators were employed to conduct the annotation task with each of them spending about 24 working months. 
To date, we have released a large-scale, high-quality, and challenging benchmark that contains over 1,000 names, 70,000 authors, and 1,000,000 papers. \figref{fig:data} shows the \sda benchmark is orders-of-magnitude larger than existing manually-labeled datasets. 

\noindent $\bullet$ \textbf{Contest leaderboard with comprehensive tasks.} 
To fairly compare various name disambiguation methods, we sponsor contests with two tracks: The first is \textit{From-scratch Name Disambiguation} (SND) aiming at grouping papers by the same author together in order to fulfill the need to create an original academic system from scratch. 
The other is \textit{Real-time Name Disambiguation} (RND), also known as incremental name disambiguation, which targets at assigning newly-arrived papers to the existing clarified authors. The RND task is crucial to maintain a regular assignment of papers on existing online academic systems owning a substantial amount of clarified author profiles. 
Beyond these, we additionally define \textit{Incorrect Assignment Detection} (IND), which attempts to remedy online paper-author affiliation errors in order to guarantee the reliability of academic systems. To date, three-round contests have been held on the first two tasks, attracting more than 3,000 researchers. Furthermore, we host a regular leaderboard to keep track of recent advances. The contest for the IND task is under active preparation.

\noindent $\bullet$ \textbf{Easy-to-use toolkit.} 
To facilitate researchers to quickly get started in the name disambiguation area, we summarize our research findings and organize an end-to-end pipeline to standardize the entire name disambiguation process, including data loading, feature creation, model construction, and evaluation,
We thoroughly investigate the contest winner methods, assemble the most effective features and models, and encapsulate them into the toolkit. The end users are free to directly invoke the baselines and encapsulated features to develop their own algorithms. 

We provide in-depth analyses of the features adopted in methods of contest winners, finding that blending the multi-modal features, i.e., the semantic features involving paper attributes and the relational features created by co-author, co-organization, and co-venue links, contributes the most to the performance of name disambiguation methods. 
On top of these discoveries, we provide simple yet effective baselines (RND/SND-all) that perform on par with the top contest methods. Particularly, RND-all has been deployed on AMiner for daily arXiv paper assignment. 


To sum up, \sda is an ongoing, community-driven, open-source project. We intend to update the leaderboard as well as offer new datasets and methods over time. We also encourage contributions at \href{mailto:oagwhoiswho@gmail.com}{oagwhoiswho@gmail.com}.


\section{\sda Benchmark}
\label{sec:dataCon}
This section first introduces the interactive annotation process for constructing the large-scale high-quality benchmark and then presents the intrinsic distributions of the benchmark.

\subsection{Interactive Benchmark Construction}
We formalize the interactive benchmark construction pipeline into two sub-modules: data collection and data annotation.

\subsubsection{Dataset Collection}
Practically, we collect the raw data from AMiner~\cite{tang2008arnetminer}. To acquire name disambiguation data with less noise and also higher ambiguity, we adopt the following rules, 


\noindent \textbf{Select authors by H-index.} For each author in AMiner, we compute the H-index~\cite{hirsch2005index}, a metric used to measure the impact of experts, and then we keep the authors with the higher H-index scores. If authors are more well-known, it is assumed that their profiles contain less noise, because they may have already clarified themselves on the academic platform. Concretely, we filtered out authors with an H-index less than 5 by sorting them in descending order based on their H-index values. This threshold is a widely accepted criterion in the literature for identifying authors with significant impact in their research field.

\noindent \textbf{Choose names with high ambiguity.} We count the number of authors with the same name in AMiner. The term ``same name'' refers to the name-blocking ways to unify names, such as moving the last name to the first or preserving all name initials but the last name~\cite{backes2018impact,kim2018evaluating}. For example, the variants of ``Jing Zhang'' include ``Zhang Jing'', ``J Zhang'' and ``Z Jing''. A name is more ambiguous if it is used by more authors. We filter names with fewer authors than a threshold to make \sda challenging\footnote{We set the threshold as 6 in \da.}.




After obtaining names with high ambiguity with the corresponding authors for each name, we collect papers for each author. Specifically, we collect the title, author names, organizations of all authors, keywords, abstract, publication year, and venue (conference or journal) as attributes of papers. Additionally, there are a large number of papers that have yet to be assigned to any authors. To increase the challenge of the benchmark, we also gather these papers, denoted as unassigned papers, whose authors share the same name as these in the benchmark, which may be assigned to the authors in the benchmark during the data annotation pipeline.


\begin{table}
	\newcolumntype{?}{!{\vrule width 1pt}}
	\newcolumntype{C}{>{\centering\arraybackslash}m{23em}}
	\caption{
		\label{tb:annotate} Data annotation pipeline. \textmd{For operations performed via three annotators, major voting is applied to solve conflicts. Anno. is the abbreviation of annotators.}
	}
	\footnotesize
	\centering 
	\renewcommand\arraystretch{1.0}
	\begin{tabular}{@{~}c?@{~}*{1}{c?C}@{~}}
		\toprule
		\textbf{Steps} &\textbf{\#Anno.}& \textbf{Operations}
		\\
		\midrule
		\makecell[c]{\textbf{Clean}\\(Roughly)}& 1
		&
		\tabincell{l}{
		 \multicolumn{1}{m{23em}}{
	    \textbf{1}. Delete papers not belonging to the concerned author;} \\
	     \multicolumn{1}{m{23em}}{
        \textbf{2}. Split over-merged author profiles into multiple authors.}
		}
		\\
		\midrule
		\textbf{Validate} & 3
		& 
		\tabincell{l}{
		\multicolumn{1}{m{23em}}{
		\textbf{1}. Same as the “Clean” step to deal with more difficult incorrect papers. 
		} } \\
		\midrule
		\textbf{Add} & 3
		& 
		\tabincell{l}{
		\multicolumn{1}{m{23em}}{
		\textbf{1}. Add unassigned papers to certain authors. 
		}
		}\\
		\midrule
		\textbf{Merge} & 3
		&  
		\tabincell{l}{
		\multicolumn{1}{m{23em}}{
		\textbf{1}. Merge separate author profiles into a single author. 
		}}  
		\\
		\bottomrule
	\end{tabular}
	
\end{table}

\subsubsection{Dataset Annotation}
\figref{fig:whoiswho_demo} demonstrates some real-world hard cases of name disambiguation, which are quite challenging for annotators to label because of the intricate relationships between papers.
In light of this, we design an interactive annotation tool\footnote{https://www.aminer.cn/billboard/id:5e42777f530c70f19522863e} adapted from ~\cite{shen2016nameclarifier} to not only provide detailed information about papers and authors  but also to offer various practical atomic operations to help annotators in performing arbitrary actions. A toy example is shown in \figref{fig:vis_tool}. The tool allows annotators to annotate interactively because each time an action is taken, the author profiles are updated and displayed to the annotators.

With the help of the tool, we establish four standardized annotation steps (detailed in Table~\ref{tb:annotate}) to ensure the manual labeling process can be conducted in a reasonable manner. Overall, the annotators are authorized to remove incorrect papers, add unassigned papers, split an author into two authors, and merge two authors. 
Specifically, the first ``\textbf{Clean}'' step allows annotators to remove or split obviously incorrect papers from the concerned author. Such papers cover different topics with the concerned author. Then, the ``\textbf{Validate}'' step allows annotators to conduct the same ``Clean'' function on incorrect papers that are hard to identify. Such papers cover relevant topics to the concerned author. After that, the ``\textbf{Add}'' step enables annotators to add unassigned papers to associated authors. Finally, the ``\textbf{Merge}'' step allows annotators to blend the papers of two authors into a single author. Since the last three steps are more challenging than the first step, three annotators are requested to annotate the same name with their results aggregated by majority voting.
Notably, annotators label all the papers of authors under the same name together each time. 
To prevent them from simply removing arbitrary papers, annotators must retain at least 80\% of the papers for each author.

In summary, 
on one hand, the devised interactive annotation process, which provides abundant facts among papers, 
fully supports annotators to label the dataset effectively.
On the other hand, each paper is examined by at least 10 skilled annotators, which further guarantees the quality of \da.



\begin{figure}[t]
	\centering
	\subfigure[Publication date]{\label{subfig:pub_date}
		\includegraphics[width=0.23\textwidth]{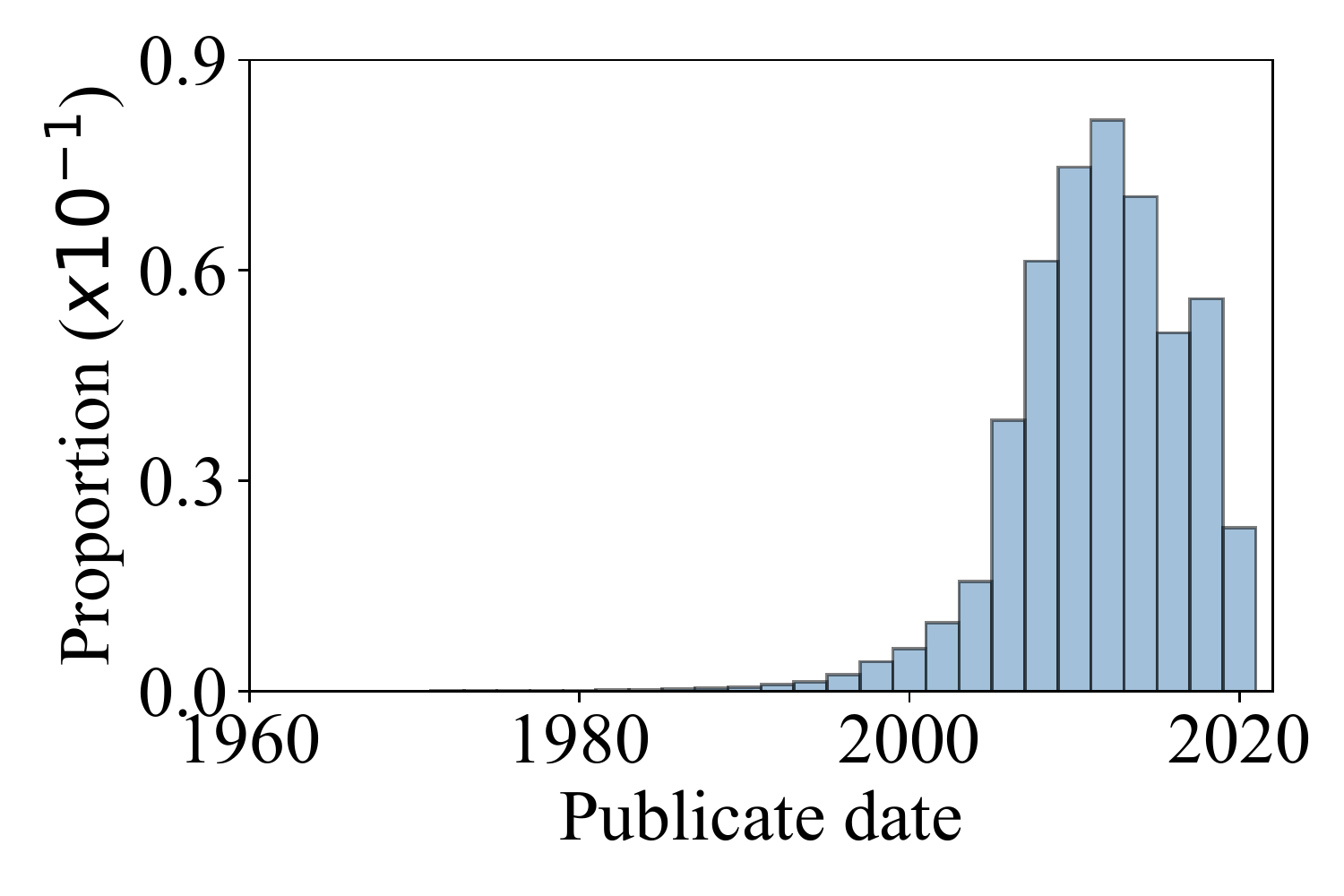}
	}
	\hspace{-0.1in}
	\subfigure[Author position]{\label{subfig:posi}
		\includegraphics[width=0.23\textwidth]{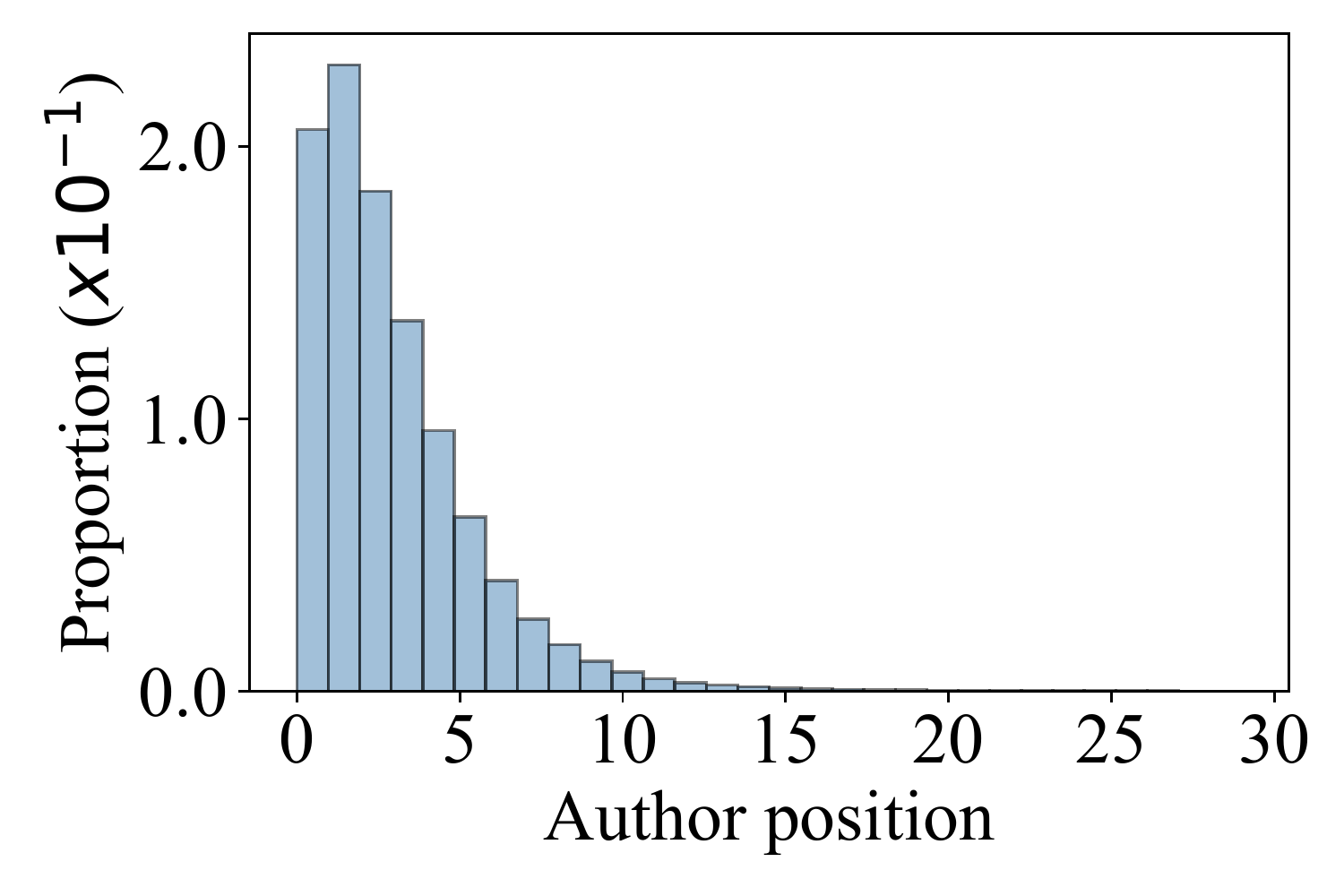}
	
	\hspace{-0.1in}}
	\subfigure[Name ambiguity (Chinese)]{\label{subfig:ch_name}
		\includegraphics[width=0.23\textwidth]{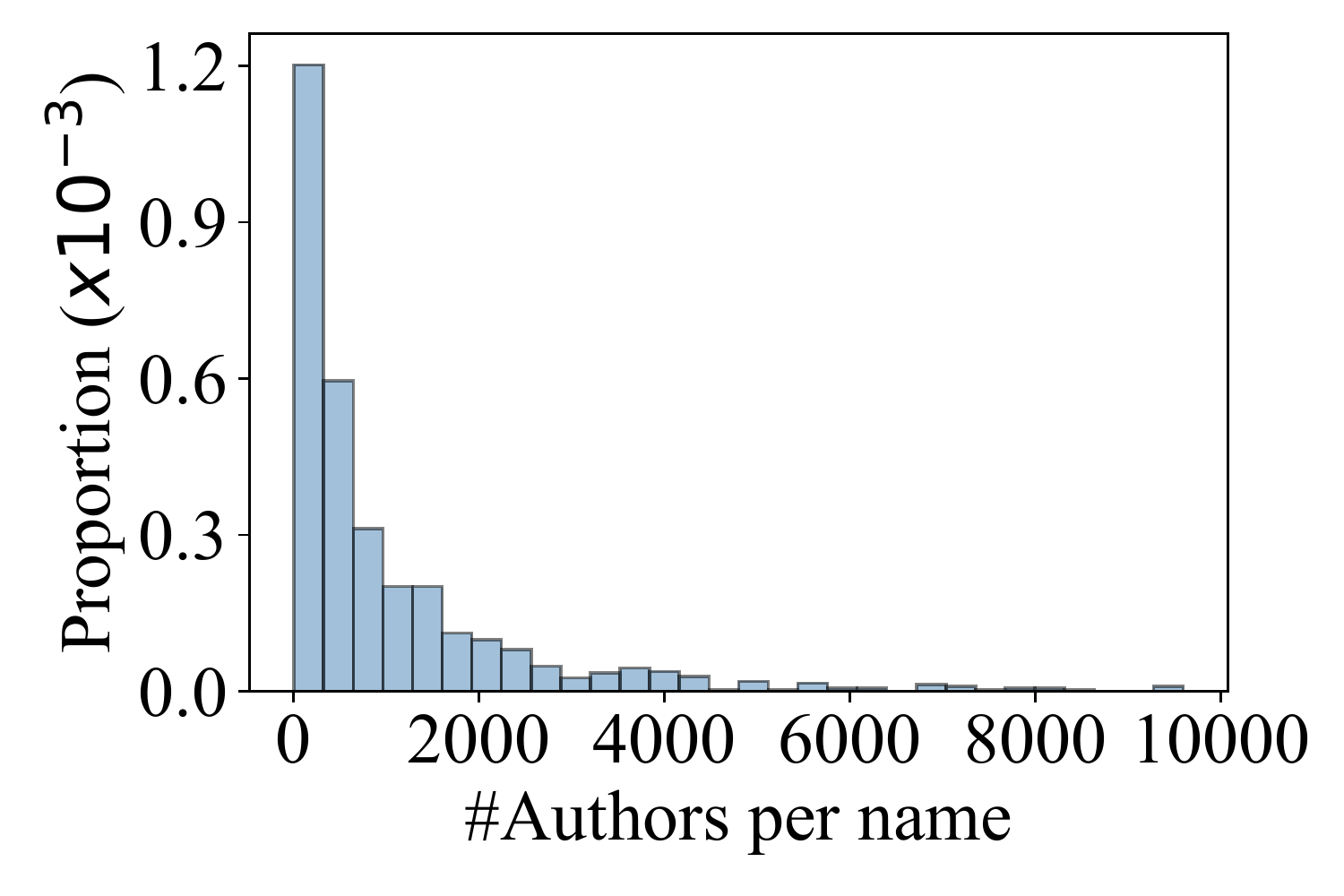}
	}
	\hspace{-0.1in}
	\subfigure[Name ambiguity (International)]{\label{subfig:for_name}
		\includegraphics[width=0.23\textwidth]{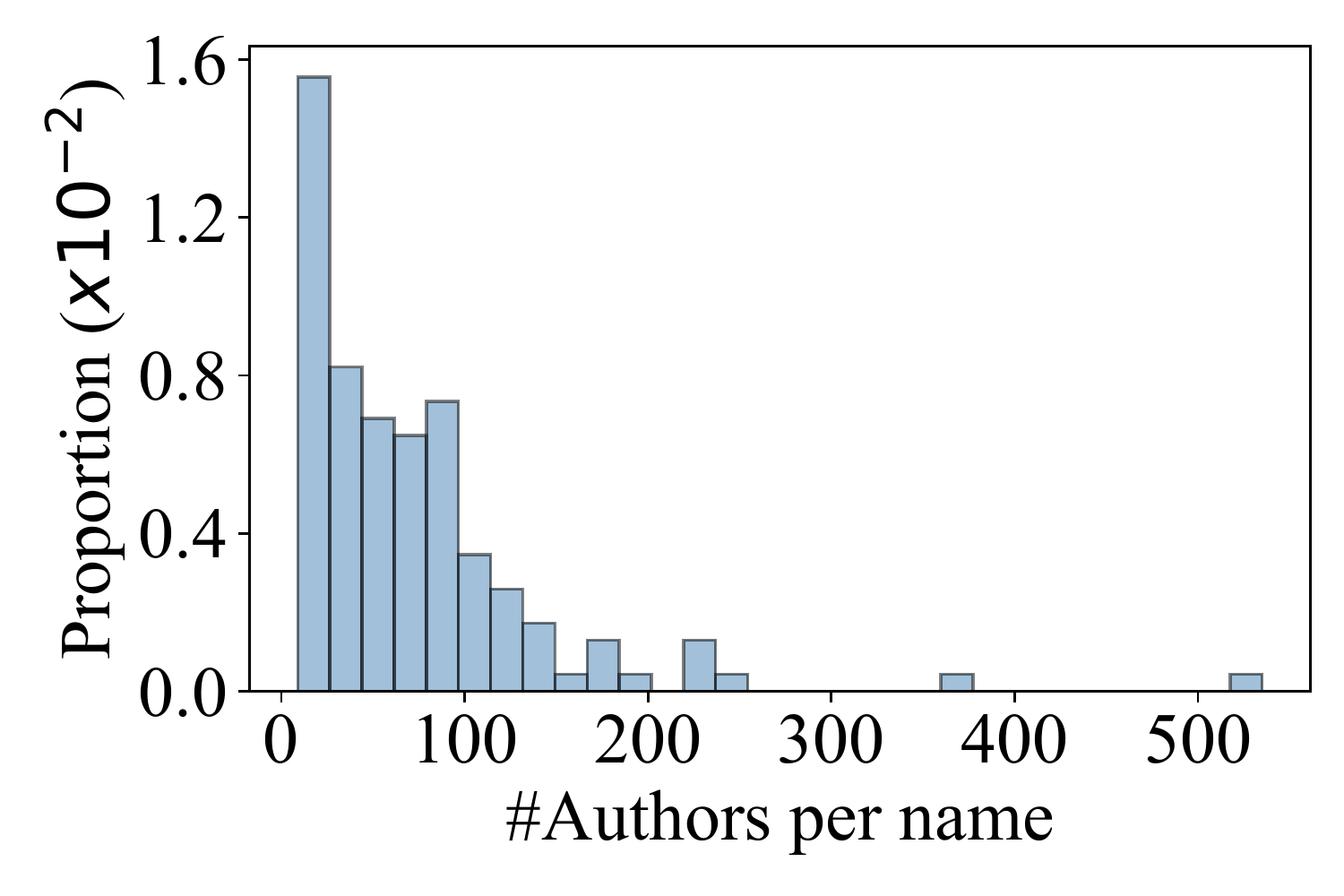}
	}
	\hspace{-0.1in}
	\subfigure[Paper number]{\label{subfig:paper_per_author}
		\includegraphics[width=0.23\textwidth]{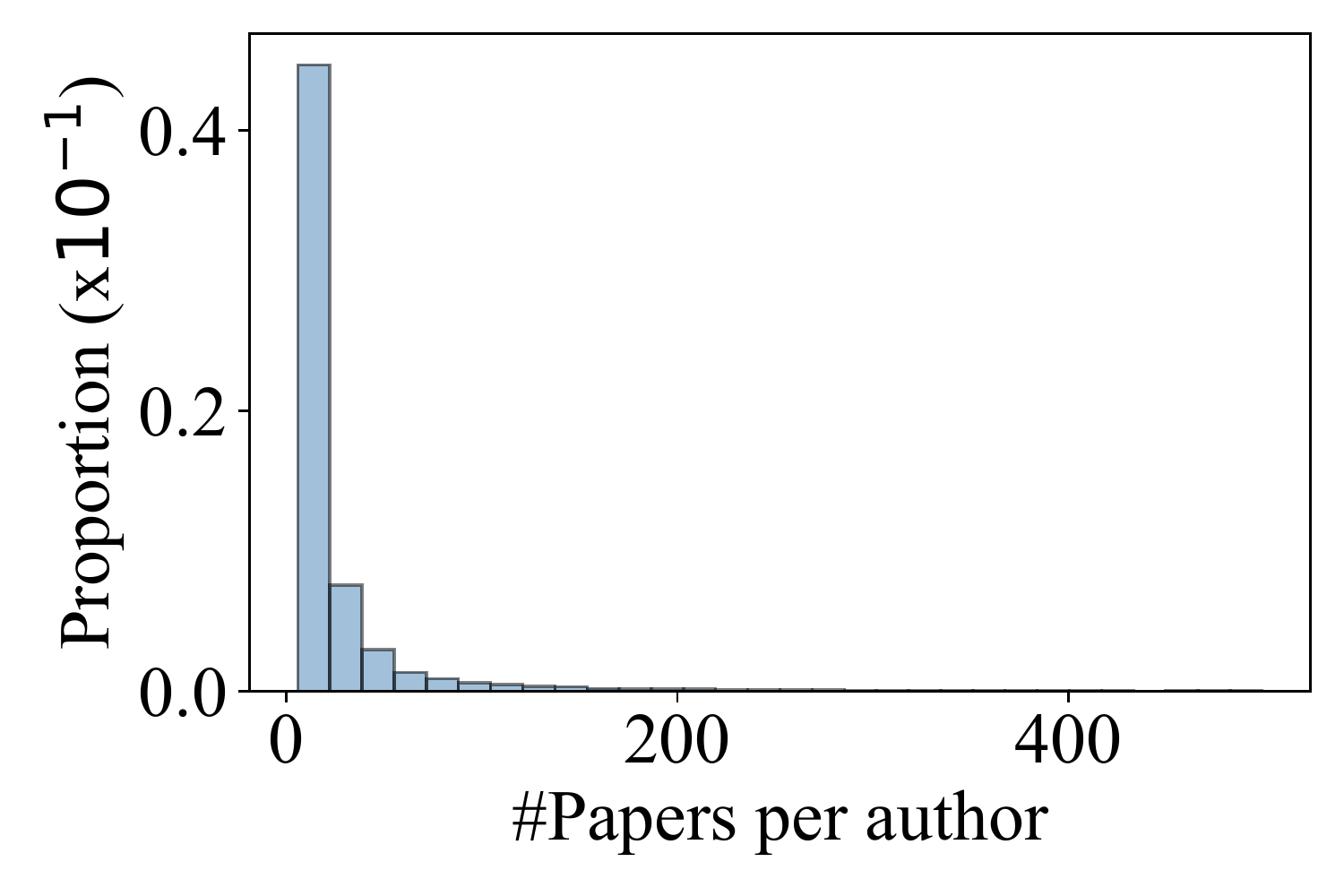}
	}
	\hspace{-0.1in}
	\subfigure[Domains (Top-10)]{\label{subfig:domain_top}
		\includegraphics[width=0.23\textwidth]{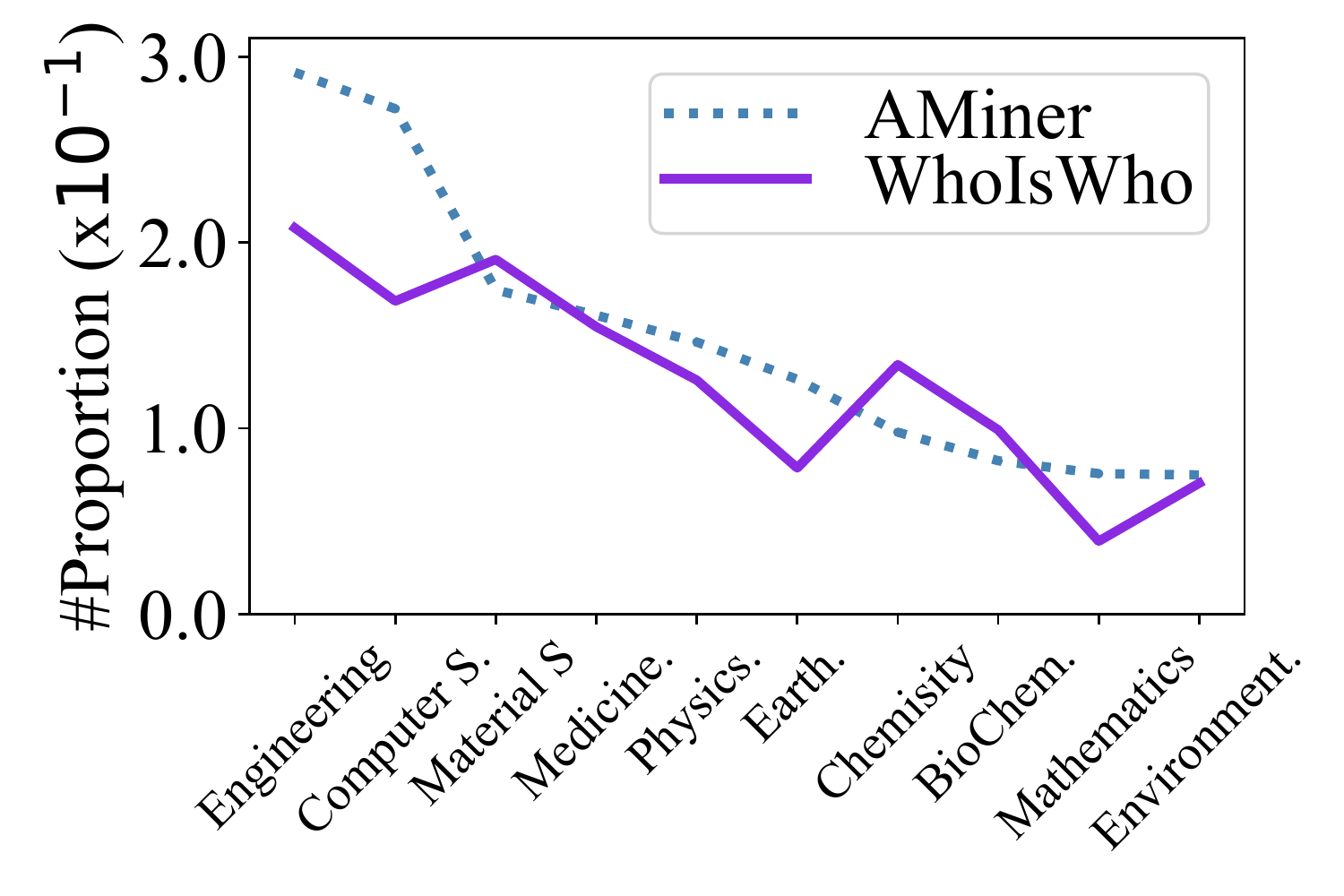}
	}
	\vspace{-10pt}
	\caption{\label{fig:datasets} Statistics of \sda benchmark}
	\vspace{-15pt}
\end{figure}

\subsection{Statistics of \sda Benchmark}
We present the holistic analysis to demonstrate the superiority of the \sda benchmark in multi-facets, as illustrated in \figref{fig:datasets}.

\vpara{Accuracy of the Annotated Authorship.} We first check the accuracy of the manually-labeled authorship. To achieve this, we randomly sample 1,000 papers from the benchmark and manually verify which papers belong to which authors. Each paper is verified by three skilled annotators via major voting.
The resultant accuracy is 99.6\% with only four assignment errors, indicating that the benchmark offers a large number of high-quantity instances.

\vpara{Publication Date Distribution.} \figref{subfig:pub_date} illustrates the distribution of paper publication date.
Few scientific documents were recorded before the year 2000 since managing digital libraries was still a relatively new technique at that time. 
As the internet develops rapidly after 2020, the number of digital records increases more quickly. 
However, there are fewer records around 2022 than there were around 2010, suggesting that the online name disambiguation system may not be able to assign the latest papers in time. 

\vpara{Author Position Distribution.} Several datasets focus on disambiguating the author on a particular position in the paper. For example, Song-PubMed~\cite{song2015exploring} is created for disambiguating the first author, which introduces biased information to  name disambiguation methods toward certain specific author positions. On the contrary, the \sda benchmark takes all author positions equally into account, as shown by the rational long-tail curve in \figref{subfig:posi}.

\vpara{Name Ambiguity Distribution.} Author names of different ethnic groups typically have varying degrees of ambiguity. Chinese authors, for example, are more difficult to disambiguate than other nationalities~\cite{gomide2017name, kim2019generating, kim2016distortive}. 
\figref{subfig:ch_name} and \ref{subfig:for_name} illustrate the distribution of the clarified author profiles per Chinese and international name respectively in AMiner, indicating Chinese names are more ambiguous than international ones.
As we focus on constructing a benchmark with high ambiguity that facilitates name disambiguation methods, we collect more Chinese names, covering about 87\% of author names in our dataset, than international names. 

\vpara{Paper Number Distribution.}
We also present the distribution of the number of papers per author in the benchmark, as shown in \figref{subfig:paper_per_author}. 
The long-tail distributions indicate that most of the cases have a  manageable quantity and only a few famous scientists own hundreds of publications. 

\vpara{Domain Distribution.}
Compared with several datasets that merely cover biased domains, such as datasets based on PubMed~\cite{song2015exploring, zhang2020mining} focus on the field of medical science, \sda has great coverage of general disciplines. To confirm
this, we randomly sample 100,000 papers and then adopt the taxonomy rank of SCImago Journal Rank (SJR)\footnote{http://www.scimagojr.com} from Scopus to obtain paper domains. The top-10 highest frequency domains are shown in \figref{subfig:domain_top}, which implies the benchmark not only covers a variety of domains but also is a representative of the overall distribution in AMiner.

\begin{figure}[t]
	\centering
	\includegraphics[width=0.48\textwidth]{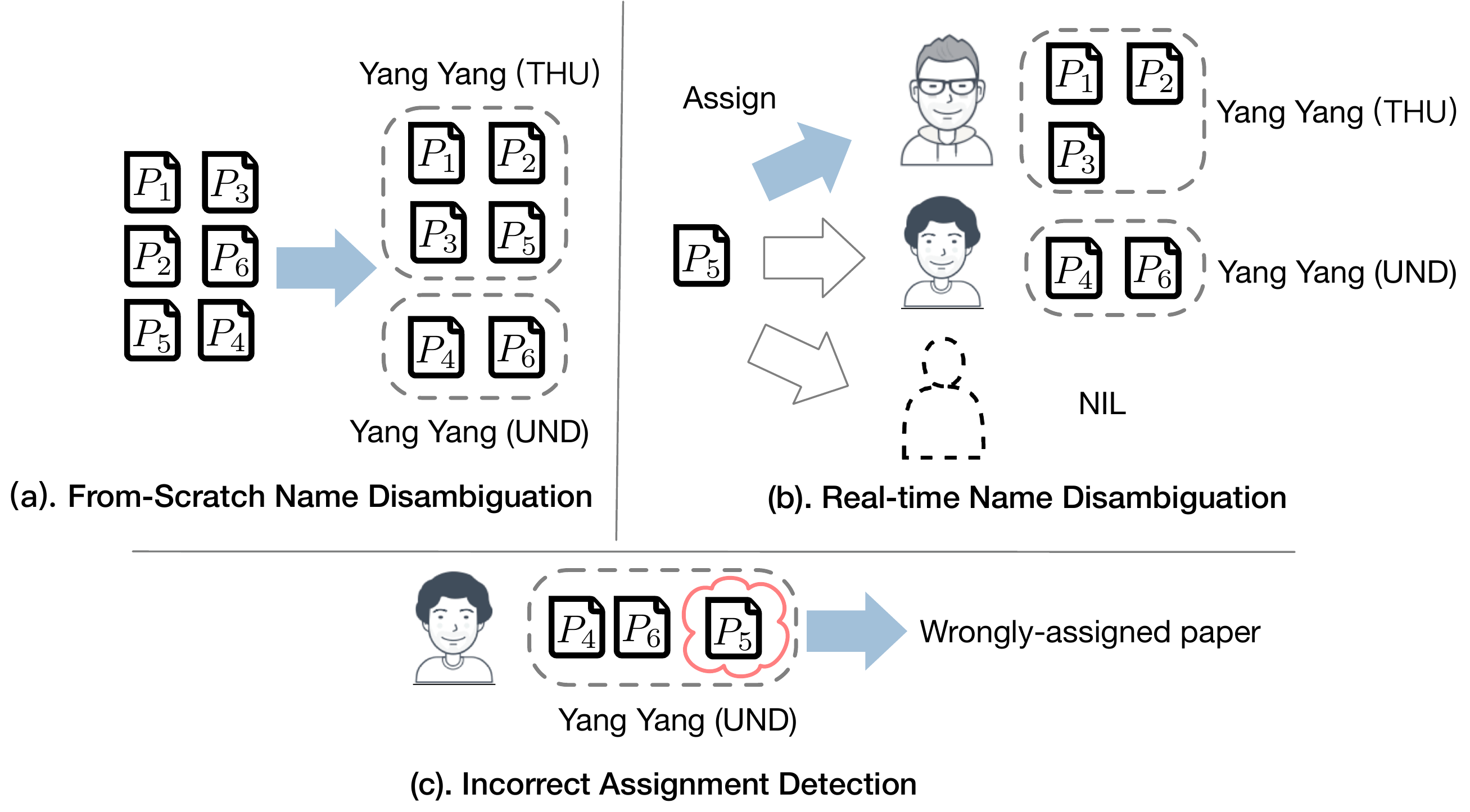}
	\vspace{-15pt}
	\caption{\label{fig:task} Three name disambiguation tasks.}
    \vspace{-10pt}
\end{figure}

\section{\sda Tasks \& Contests}
In this section, we first present three name disambiguation tasks with standardized evaluation protocols. Then we review three-round historical contests and a regular leaderboard  built on defined tasks with different released versions. 

\subsection{Task Formations and Evaluation Protocols}
Here we formalize the three tasks
i.e, from-scratch name disambiguation, real-time name disambiguation, and incorrect assignment detection, 
with evaluation metrics, as shown in \figref{fig:task}.  
\begin{definition}
	\textbf{Paper}. A paper $p$ is associated with multiple fields of attributes, i.e., $p = \{x_1, \cdots, x_F\}$, where $x_f \in p$ represents the $f$-th attribute. 
	$F$ is the number of attributes.
\end{definition}	

\begin{definition}
	\textbf{Author}. An author $a$ is comprised of a set of papers, i.e., $a=\{p_1, \cdots, p_{n}\}$, where each paper $p_i=\{x_1, \cdots, x_F\}$ and $n$ is the number of papers authored by $a$. 
\end{definition}

\begin{definition}
	\textbf{Candidate Papers}. Given a person name denoted by $na$, $\mathcal{P}^{na} = \{p^{na}_1, \dots, p^{na}_N\}$ is a set of candidate papers written by any author with the name $na$.
\end{definition}

\begin{definition}
	\textbf{Candidate Authors}. Given a person name denoted by $na$, $\mathcal{A}^{na} = \{a^{na}_1, \dots, a^{na}_M\}$ is a set of candidate authors with the same name $na$. The term ``same name'' refers to the ways to unify names using name blocking techniques~\cite{backes2018impact,kim2018evaluating}.
\end{definition}

\subsubsection{From-scratch Name Disambiguation}
At the beginning of building digital libraries, we need to partition a large number of published papers into groups, each of which represents papers that belong to a single person. To achieve this, we formalize from-scratch name disambiguation as a clustering problem.



\begin{problem}
	\textbf{From-scratch Name Disambiguation (SND)}. Given a set of candidate papers $\mathcal{P}^{na}$,
SND aims at finding a function $\Phi$ to partition $\mathcal{P}^{na}$ into a set of disjoint clusters $C^{na}$, i.e.,
	
	\beq{
		\Phi(\mathcal{P}^{na}) \rightarrow C^{na}, \text{where } C^{na}=\{C^{na}_1, C^{na}_2, \cdots, C^{na}_K\}, \nonumber
	}

\noindent where each cluster consists of papers owned by the same author, i.e., $\mathbb{I}(p_i^{na})=\mathbb{I}(p_j^{na}), \forall (p_i^{na}, p_j^{na}) \in C_k^{na}\times C_k^{na}$,  and different clusters 
contain papers from different authors, i.e., $\mathbb{I}(p_i^a)\ne \mathbb{I}(p_j^a), \forall  (p_i^a, p_j^a) \in C_k^a\times C_{k'}^a, k \ne k'$. $\mathbb{I}(p_i^{na})$ is the author identification of the paper $p_i^{na}$.
\end{problem}

\vpara{Evaluation Protocol.} We adopt the macro pairwise-F1 to evaluate the performance of related SNA methods, which is widely adopted by many SND methods~\cite{zhang2018name, subramanian2021s2and, zheng2021dual, santini2022knowledge, li2021disambiguating}. 




\subsubsection{Real-time Name Disambiguation}
Assigning new papers to existing authors is crucial for online digital libraries at the current stage.
For instance, AMiner receives over 500,000 new papers each month. 
To this end, we formalize the real-time name disambiguation as a classification problem.


\begin{problem}
	\textbf{Real-time Name Disambiguation (RND).} Given a paper $p^{na}$, i.e, the paper with its author name $na$ to be disambiguated, and the set of candidate authors $\mathcal{A}^{na}$, the right author $a^*$ can be either a real author in $\mathcal{A}^{na}$ or a non-existing author profile, i.e., NIL. 
	We target at learning a function to assign the paper $p^{na}$ to $a^*$, i.e., 
	\beq{
	 \Psi (p^{na}, A^{na}) \rightarrow  a^*  \nonumber
	} 
\end{problem}
Note that NIL situations are found frequently in online academic platforms. 
Assuming that undergraduate students publish their first paper at a conference or journal, but the current database has not yet established their author profile, 
it is infeasible to assign the paper to any authors. 
In light of this, we have incorporated the NULL scenarios in the RND task. Formal efforts~\cite{chen2020conna} also take into account the NIL situation, however, they create synthesized NIL labels rather than incorporating the actual NIL cases.
To our best knowledge, we are the first to consider the NIL situation in the \sda benchmark with manually-labeled real NIL cases.


\vpara{Evaluation Protocol.} We propose the weighted-F1 to evaluate the methods that solve the RND problem. For an author $a$ to be disambiguated, we calculate the metrics as follows:

\begin{equation}
\begin{aligned}
    Precision^{a} &= \frac{\#PapersCorrectlyAssignedToTheAuthor}{\#PapersAssingedToTheAuthor}, \\
    Recall^{a} &= \frac{\#PapersCorrectlyAssignedToTheAuthor}{\#PapersOfTheAuthorTobeAssinged}, \\
    Weight^{a} &=  \frac{\#PapersOfTheAuthorTobeAssigned}{\#TotalPapersTobeAssigned},
    \nonumber
\end{aligned}
\end{equation}

\noindent where precision measures the correctness of papers predicted to $a$, and recall measures how many papers from $a$'s actual papers could be correctly assigned to $a$. Then we calculate the F1 score by the precision and recall for each author. After that, we average the F1 score by the weight of each author which is determined by the percentage of their papers that will be assigned. 
We adopt the weighted average strategy to alleviate the negative effects of some extreme cases, like authors who only have one paper.

\subsubsection{Incorrect Assignment Detection} 
As inevitable cumulative errors brought via the methods of SND and RND greatly affect the efficacy of subsequent assignments, Incorrect Assignment Detection is a vital task to detect and remove wrongly-assigned papers.

	

\begin{problem}
\textbf{Incorrect Assignment Detection (IND).} Given a conflated author entity $a^* = \{p_i, \cdots, p_j, p_a, \cdots, p_b, \cdots, p_m, \cdots, p_n\}$ comprising multiple papers from $K$ different authors $\{a_1, \cdots, a_K\}$, where $a_1 = \{p_i, \cdots, p_j\}$, $a_2 = \{p_a, \cdots, p_b\}$, and $a_K = \{p_m, \cdots, p_n\}$. Assuming $a_1$ covers the highest percentage of papers within $a^*$, we set $a^* = a_1$. Consequently, the papers owned by $\{a_2, \cdots, a_K\}$ are defined as incorrectly-assigned papers to be detected.
\end{problem}

\vpara{Evaluation Protocol.} We leverage Area Under ROC Curve (AUC), broadly adopted in anomaly detection~\cite{ma2021comprehensive} and Mean Average Precision (MAP), which pays more attention to the rankings of incorrect cases, as the evaluation metrics.

\subsubsection{Discussion}
The proposed three name disambiguation tasks shed light on the life cycle of concerned name disambiguation problems in online digital libraries. Specifically, the SND task reflects the requirements of building digital libraries at the early stage; 
the RND task corresponds to the urgent needs of current online platforms;
and the IND task is devoted to correcting the accumulated errors of name disambiguation algorithms, which is critical to maintaining the reliability of the name disambiguation system. 
In addition, the three tasks can serve as the backbone of any other complex name disambiguation tasks. 
We believe name disambiguation methods, which perform better on these tasks, are powerful enough to handle the majority of name disambiguation situations. Although \citet{zhang2021name} have already proposed similar types of tasks, we improve them by 1) taking the NIL issue into account and formalizing the RND problem into a more general classification problem instead of a ranking problem, 2) standardizing the evaluation protocol of the three tasks, and 3) arranging contests for the first two tasks to prompt their accomplishments.


\subsection{Historical Contests \& Regular Leaderboard}
 \label{sec:comp}
\begin{figure}[t]
	\centering
	\includegraphics[width=0.48\textwidth]{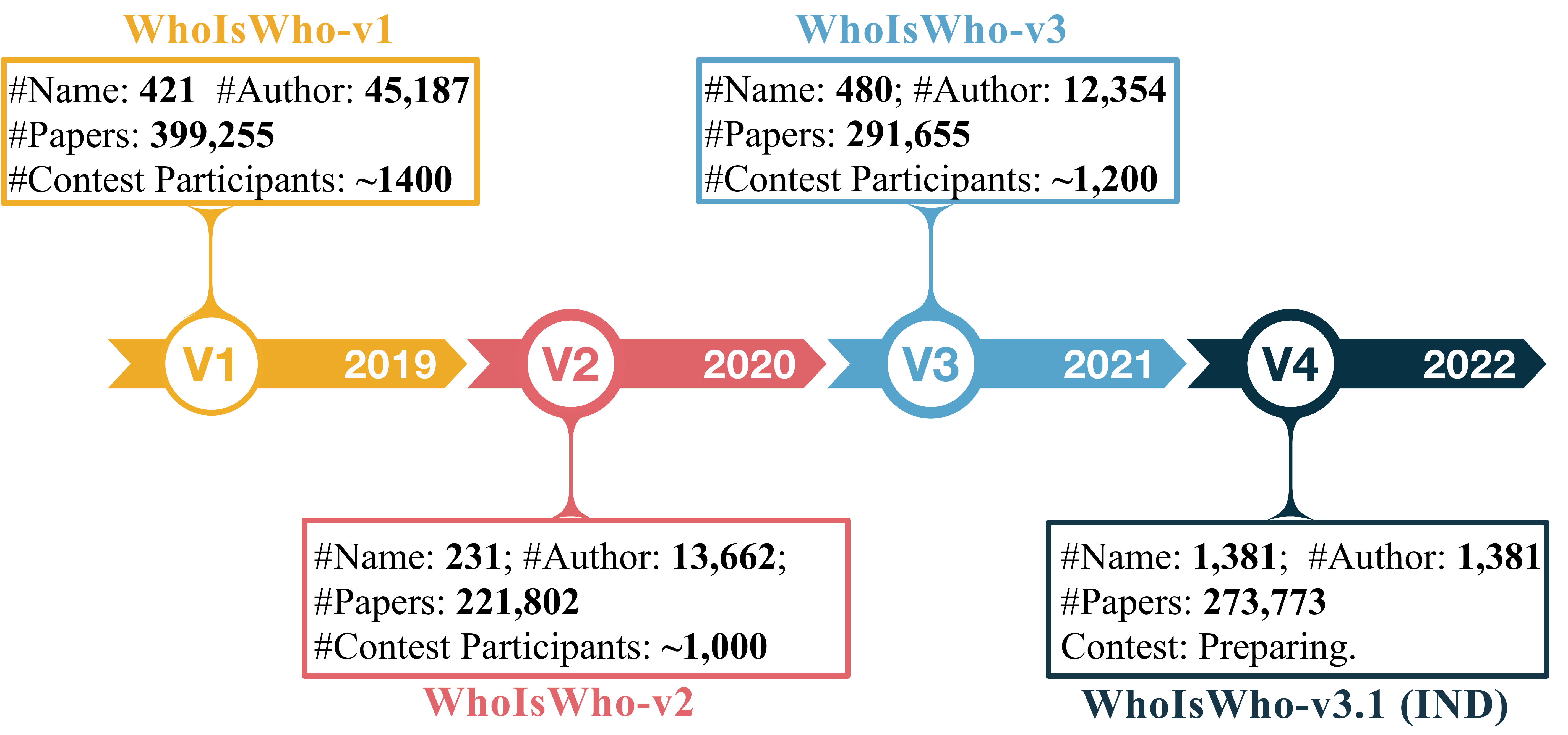}
	\vspace{-10pt}
	\caption{\label{fig:timeline} The released time of \sda benchmark and launched contests.}
	\vspace{-15pt}
\end{figure}

 From 2019 to 2022, \sda periodically released three versions of benchmarks. 
 To promote the development of the community, we sponsored three rounds of name disambiguation contests on BienData\footnote{https://www.biendata.xyz/}. 
The timeline of released benchmarks and corresponding contests is depicted in \figref{fig:timeline}.
To date, more than 3,000 people in the world, 
have downloaded the \sda benchmark more than 10,000 times. \sda has already become one of the most well-known and representative benchmarks of the name disambiguation community. In addition, to assist researchers who are interested in resolving name disambiguation problems at any time, we maintain a regular leaderboard with the contest based on the most recent benchmarks released by \da. 



In the following part, we briefly revisit the methodologies proposed by contest winners, based on which we conduct an in-depth empirical analysis to probe key factors that may have a significant impact on the performance of name disambiguation methods.

\subsubsection{Methodologies of the Contest Winner}
We revisit the approaches of contest winners in the first two tasks of SND and RND since they have the best performance to date.
How to measure the fine-grained similarities between papers and authors is vital to finding a solution to both tasks. 
Thus, to measure these similarities, we need to build the interaction between authors and papers, which needs to be primarily explored. In the following part, we skip over some technical details and focus on the strategies to quantify connections between papers and authors.


\vpara{From-scratch Name Disambiguation.}
The SND task aims to group the papers written by the same author. The contest winner divides the similarities across papers into two categories.

\ipara{Semantic Aspect.} 
The contest winner views the paper's title, venue, organization of authors, year, and keywords as the semantic features, based on which they measure the topical similarities between papers. 
Specifically, they first learn word2vec~\cite{mikolov2013efficient} embeddings based on the semantic features of all the papers in the \sda benchmark.
Then they project the semantic features of a paper into 
corresponding word embeddings and average them as the paper embedding.
Finally, they calculate the soft semantic similarities between papers based on these semantic embeddings.

\ipara{Relational Aspect.} The contest winner takes author names and organizations as the relational features of papers. For example, the concurrence of the same author name in two papers 
reflects their relationships. 
Specifically, they construct a relational graph by considering papers as nodes and the connections between papers as edges. 
If two papers have identical coauthors' names, 
the edge of co-authors is added.
When two papers have the same organization for the concerned author, 
the edge of co-organization is added.
After that, they employ the metapath2vec~\cite{dong2017metapath2vec} to obtain relational embeddings of papers. Finally, they calculate the relational similarity score between papers based on these relational embeddings.



Furthermore, the contest winner combines the two multi-modal similarities to estimate the final similarities between papers and then uses DBSCAN~\cite{ester1996density} to obtain the clustering results.


\vpara{Real-time Name Disambiguation.}
The RND task focuses on measuring connections between the paper and a collection of papers from each candidate author. The contest winner captures more precise semantic features between unassigned papers and candidate authors than the SND task as follows.

\ipara{Semantic Aspect.} Besides the soft semantic features, i.e., those measured via embedding techniques, they also consider the ad-hoc semantic features, i.e., those measured via hand-crafted features. 
In terms of the soft semantic features, they identify similarities between the target paper and each paper of the candidate author, just like SND does. Then they adopt aggregation functions to obtain overall similarities between the target paper and all papers of the candidate author.
As for the ad-hoc semantic features, they propose 36-dimensional hand-crafted features to explicitly capture the semantic correlations between the target paper and the candidate author. The complete features are listed in Table~\ref{tb:handfeatures}. Finally, they concatenate the soft semantic features and the ad-hoc semantic features to create the final similarity features.
Then they adopt ensemble methods to acquire the classification results.

Being aware that the contest winner's methods disregarded the characterization of relationship properties. We make the following hypotheses:
1) Unlike the SND task, which only requires building a relational graph of papers from one name once, the RND task needs to build time-consuming graphs between unassigned papers and corresponding candidate authors with each unassigned paper once.
2) Some ad-hoc features can somewhat capture relational correlations. For example, the coauthor-occurrence feature, which counts the number of coauthors between the target paper and a candidate author, can be viewed as the coauthor edge weight on virtual paper-author graphs. 
Nevertheless, how to model the relational correlations in the RND task is still under-explored.

\vpara{Incorrect Assignment Detection.} The IND task targets at detecting accumulated incorrect papers,
which is important to guarantee the reliability of academic systems. However, there is no available IND benchmark in the current stage. To this end, we have released V3.1 data consisting of 1,000+ authors and  200,000+ papers dedicated to the IND task. To our best knowledge, we are the first to specify and release the corresponding IND benchmark. Furthermore, we are planning a contest based on the released \da-v3.1 benchmark for the IND task in a few months.




\subsubsection{Discussion}
\label{subsubsec:contest_obser}
In summary, we observe a crucial insight of establishing a good approach to comprehensively measure the correlations among papers is to intertwine multi-modal features, i.e., semantic and relational features. 
The contest results show that methods capturing both two aspects of features produce impressive results. Although the contest for the third task IND has not been held, we assume a similar result may be drawn for the IND task, as they also depend on evaluating the agreements among papers.  

\begin{figure}[t]
	\centering
	\subfigure[Soft semantic features of SND.]{\label{subfig:sna_se}
		\includegraphics[width=0.23\textwidth]{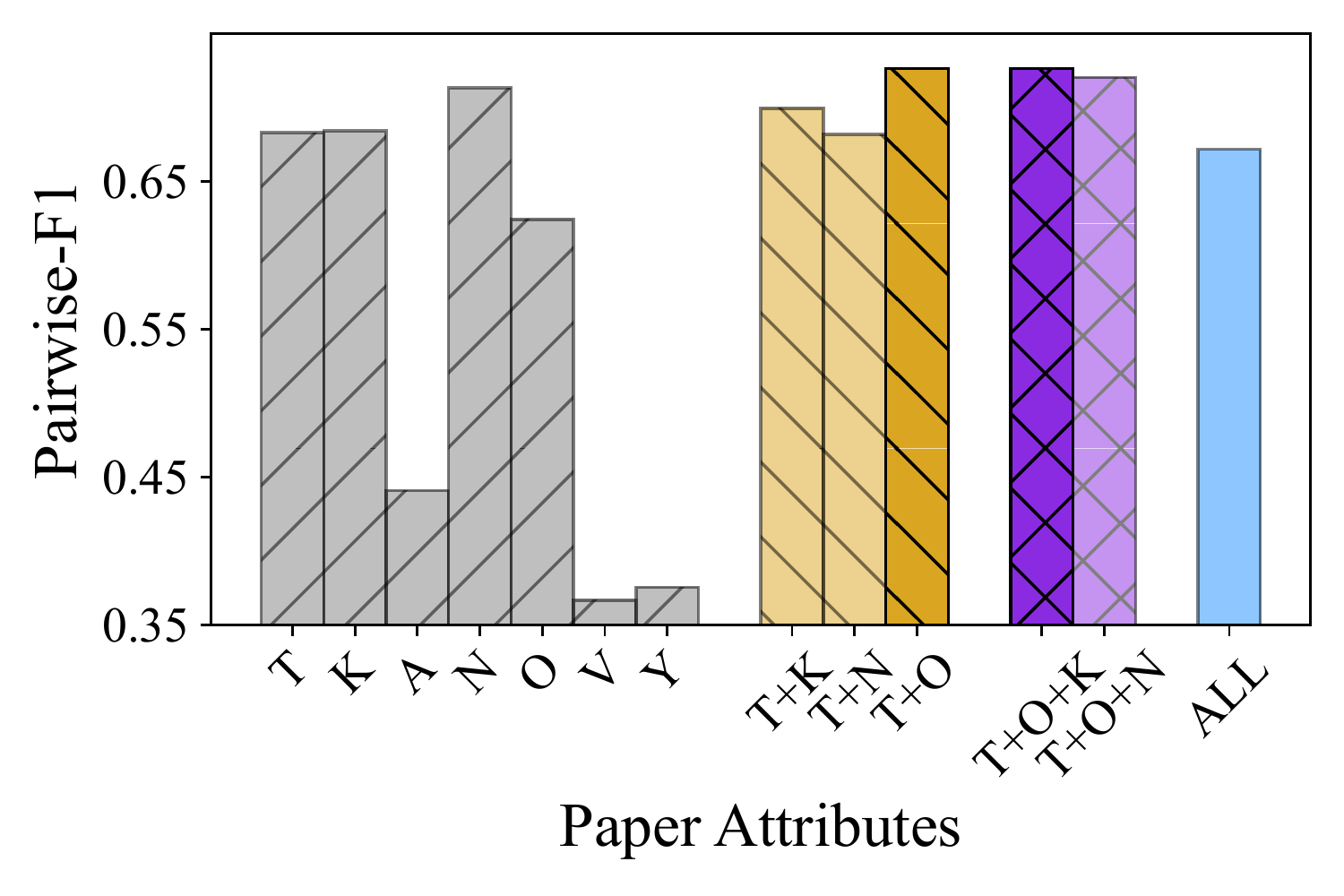}
	}
	\hspace{-0.1in}
	\subfigure[Relational features of SND.]{\label{subfig:sna_str}
		\includegraphics[width=0.23\textwidth]{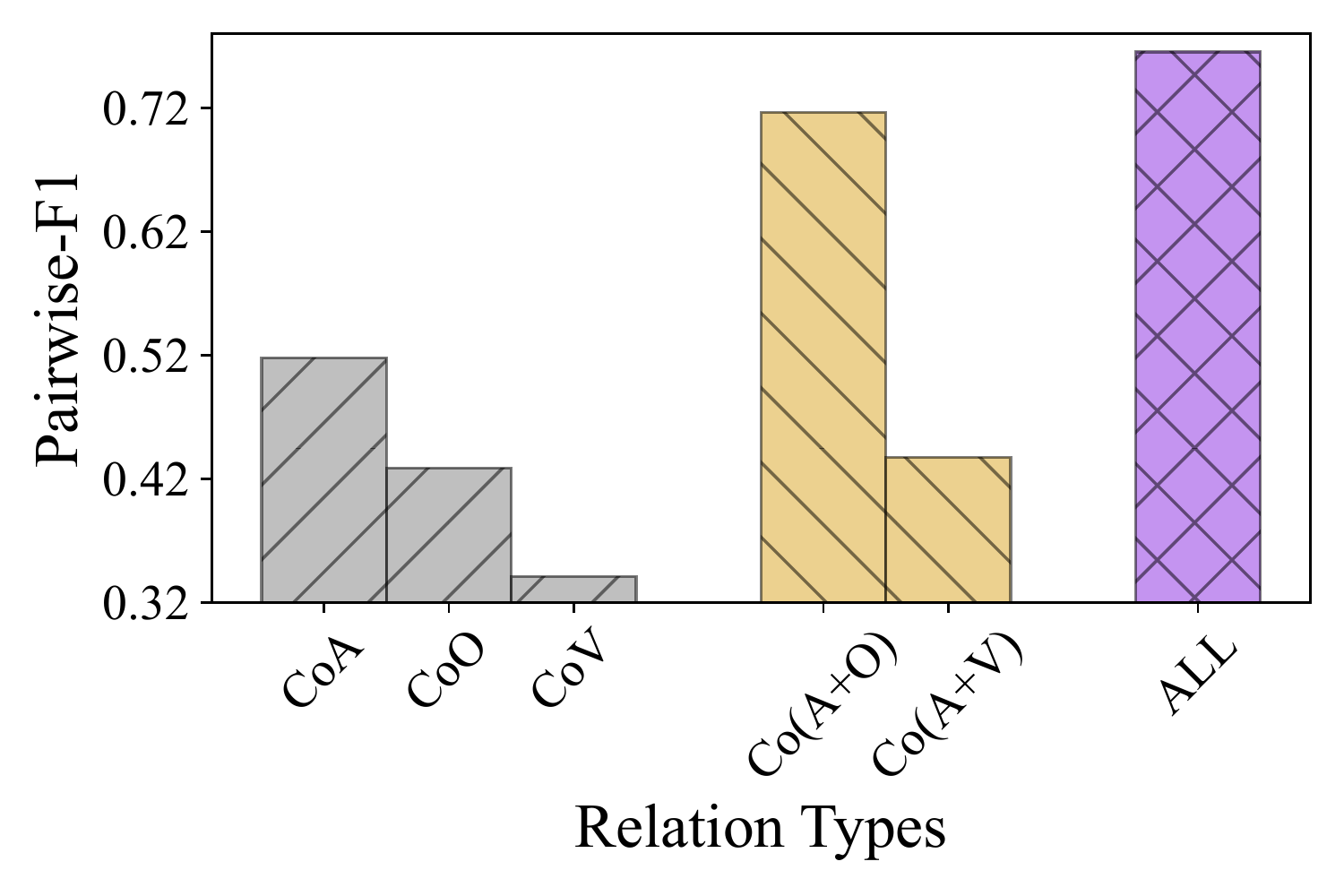}
	}
	\hspace{-0.1in}
	\subfigure[Soft semantic features of RND.]{\label{subfig:ina_soft}
		\includegraphics[width=0.23\textwidth]{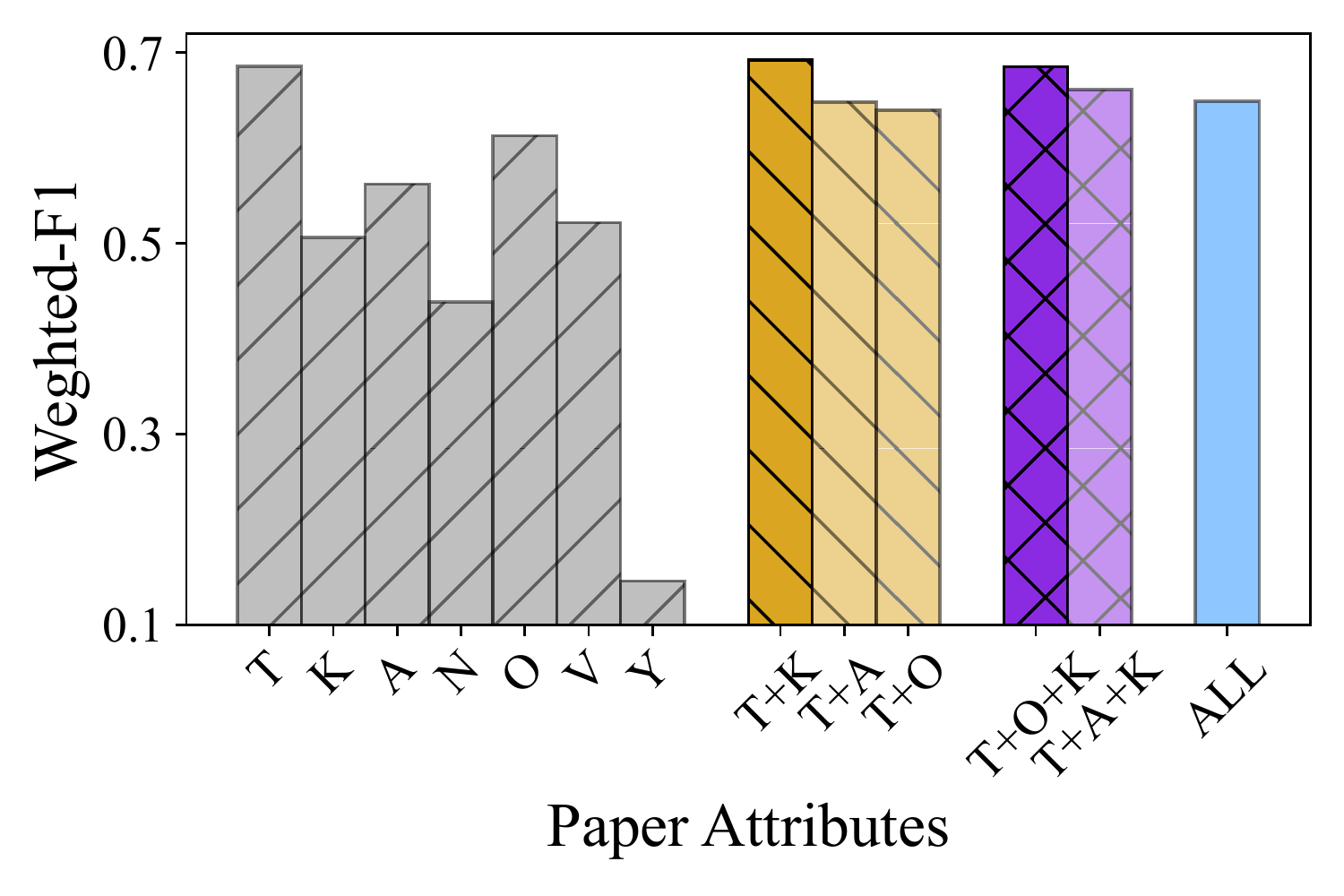}
	}
	\hspace{-0.1in}
	\subfigure[Ad-hoc semantic features of RND.]{\label{subfig:ina_hard}
		\includegraphics[width=0.23\textwidth]{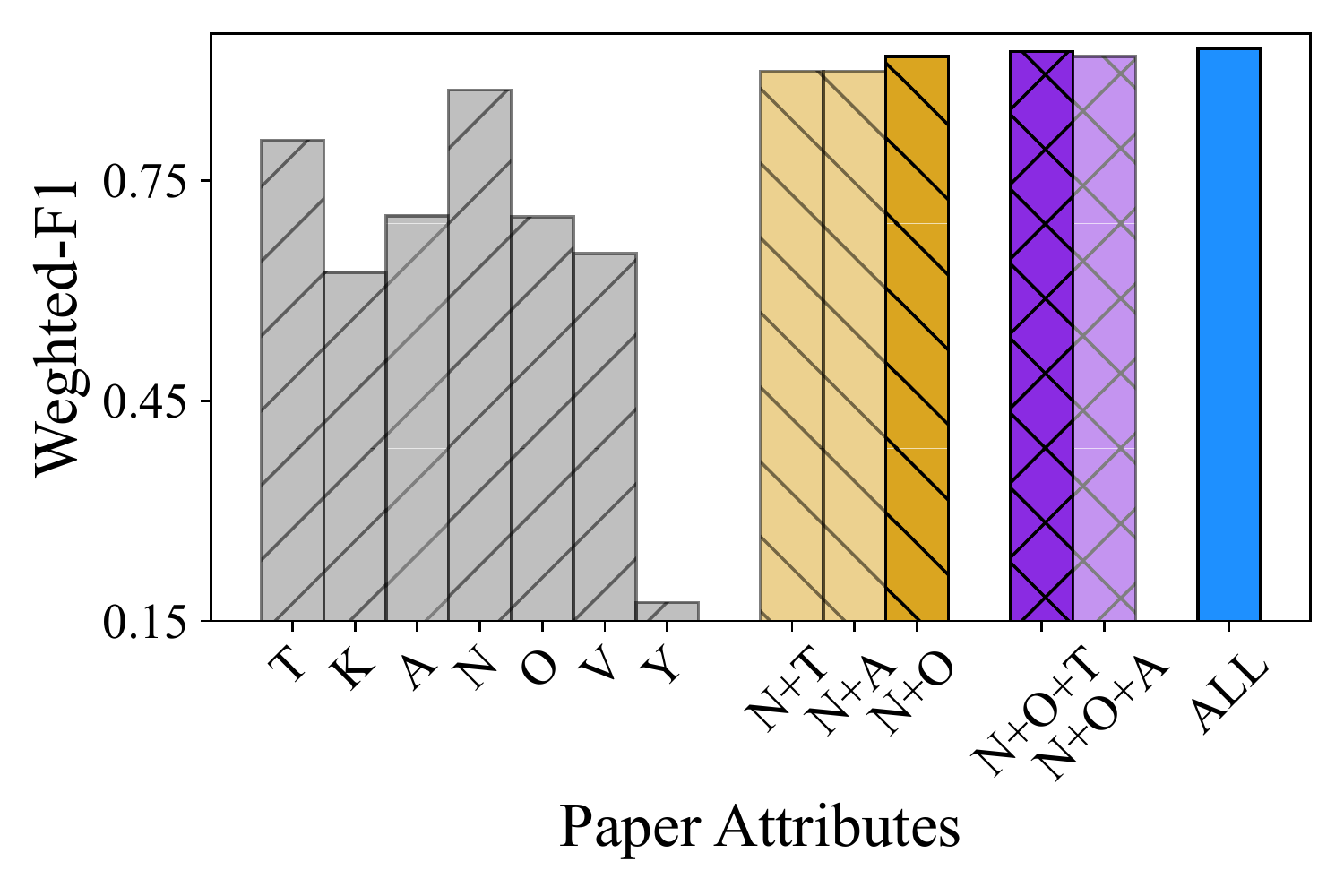}
	}
	\vspace{-5pt}
	\caption{\label{fig:feature_import} \textbf{Feature importance on the SND and RND tasks.}}
	\vspace{-10pt}
\end{figure}

\section{Empirical Factor Analysis}
\label{sec:exp}
We conduct in-depth ablation studies to understand the effect of various factors on name disambiguation performance.
To ensure fair comparisons, we only modify the factors of interest, leaving others unaltered. 
We adopt metrics defined in \sda tasks for evaluations. 
For each experiment, we run 5 trials and report the mean results at the \da-v3 validation set.



\subsection{Semantic Feature Importance}
\label{subsubsec:semantic}
We study the effects of accessible paper attributes, i.e., title (T), keywords (K), abstract (A), venue/journal (V), year (Y), author names (N), and organizations of authors (O), on the SND and RND tasks.

\vpara{From-scratch Name Disambiguation.} To perform the soft semantic feature analysis, we adopt a similar implementation pipeline with the contest winner method while exploring different attributes.

\vpara{Results.} The results are shown in Fig.~\ref{subfig:sna_se}. 
The fields of title, keywords, author name, and organization play a more significant effect on disambiguation than others. 
The field of abstract contains much redundant words and noises.
The venue and year also fail to access the similarities among papers.
\textbf{(1) Combining consistent attributes might better express semantic correlations.}
We combine these four effective single features, as shown in the yellow bars. The title + name even perform worse than its constituent single attribute. 
We speculate that compared to title, keywords, and organization which have semantic correlations among papers, the author's name has more linguistic qualities. 
Thus, combining two disparity attributes result in performance degradation.
The performance of the title improves when it is paired with keywords or organization, suggesting that a consistent attributes combination may better express semantic correlations. 
\textbf{(2) Combining title, keywords, and organization performs the best}.
Finally, the combination of title, keywords, and organization, represented by the purple bars, performs better than mixing all the attributes together, i.e., the blue bar.
This suggests that adding more attributes without calibrating may result in noise and lower performance.

\vpara{Real-time Name Disambiguation.} 
We also adopt the RND contest winner method's implementation pipeline.
In addition to the soft semantic feature analysis, we also  
explore how various paper attributes affect the performance of name disambiguation methods using hand-crafted features listed in Table~\ref{tb:handfeatures}.

\vpara{Results.} The results are shown in \figref{subfig:ina_soft} and \ref{subfig:ina_hard}. 
\textbf{
(1) The soft semantic features share a similar trend on both tasks.
}
Regarding the soft semantic features, 
\figref{subfig:ina_soft} and \ref{subfig:sna_se} show that both tasks share a common trend: 1) the attributes of title, keywords, and organization perform well and 
2) the combination of title, keywords, and organization performs better than just mixing all the considered features. 
This is expected because both tasks measure the agreements between papers and authors via the same soft semantic feature modality.
In terms of the ad-hoc semantic features, shown in \figref{subfig:ina_hard}, 
the author name is the most effective factor to determine the performance of algorithms. 
\textbf{
(2) Mixing all attributes performs best.
}
Surprisingly, the blue bar, which represents the performance of combining all features, outperforms other combination patterns, suggesting that despite falling into the semantic feature category, the ad-hoc feature characterization frameworks have different underlying biases than the soft one.




\begin{table}
	\newcolumntype{?}{!{\vrule width 1pt}}
	\newcolumntype{C}{>{\centering\arraybackslash}p{3.3em}}
	\renewcommand\tabcolsep{3.5pt} 
	\caption{
		\label{tb:modarity_res} Performance (\%) of different feature modalities (semantic or relational) and their combinations.
	}
	\footnotesize
	\centering 
	\renewcommand\arraystretch{1.0}
	\begin{tabular}{@{~}l?@{~}*{1}{CC?}*{1}{CC?}*{1}{C}@{~}}
		\toprule
		
		\multirow{2}{*}{\vspace{-0.3cm} Tasks}
		&\multicolumn{2}{c?}{\textbf{Semantic Feature}}
		&\multicolumn{2}{c?}{\textbf{Relational Feature}} 
		&\multirow{2}{*}{\textbf{All}} 
		\\
		\cmidrule{2-3} \cmidrule{4-5} 
		&\textbf{Soft}
		&\textbf{Ad-hoc}
		&\textbf{Relation}
		&\textbf{Ego}
		&
		\\
		\cmidrule{1-3} \cmidrule{4-6} 
		\textbf{SND}
		&  72.64 & - &  76.52  & - & \textbf{88.46}\\
		\textbf{RND}
		&  76.55 & 93.01 &  - & 72.92 & \textbf{93.40} \\
		\bottomrule
	\end{tabular}
\end{table}

\subsection{Relational Feature Importance}
\label{subsubsec:relation_import}
Empirically, the fields of the author name and venue show a greater relational dependency between papers. Moreover, the field of organization has both relational and semantic characteristics. Therefore, we build three relational edges between papers: 
CoAuthor, where two papers have a relationship only if they share the same author name;
CoOrg, where two papers have a relationship only if they share the same affiliations\footnote{We only take the organization of the author to be disambiguated into consideration.};
CoVenue, where two papers have a relationship only if they are published in the same venue or journal.

\vpara{From-scratch Name Disambiguation.} We also follow the implementation pipeline of the contest winner method to obtain the relational paper embeddings in the built rational graphs, while exploring the effects of different relational edges. 

\vpara{Results.} Fig.~\ref{subfig:sna_str} presents the performance of using different relation types. The grey bars, which show that CoAuthor performs the best among the single relational types, suggest that the author name has more important relational information than the semantic information. CoVenue performs the worst because massive papers from various domains may be published in the same venue/journal. 
Combining all three features yields the best results when taking into account the mixed outcomes, represented by the yellow and purple bars, which is consistent with empirical findings from \secref{subsubsec:semantic} that consistent attribute combinations can improve performance.

\subsection{Feature Modality Importance}
\label{subsec:mix_feat}
We explore how the semantic and relational features affect the effectiveness of disambiguation. We conduct a thorough examination about the combination patterns of multi-modal features to see which ones perform the best.
For the SND task, we leverage the paper attributes of title, keywords, and organization as the soft semantic features.
For the relational features, we adopt three relation types, i.e., CoAuthor, CoOrg, and CoVenue.
For the RND task, 
in addition to the soft and ad-hoc semantic features used in \secref{subsubsec:semantic}, 
we build the heterogeneous ego-graph for each pair of the target paper and a candidate author in order to add relational features. 

\vpara{Results.} Table~\ref{tb:modarity_res} shows the performance of single feature modalities and their combinations. 
\textbf{(1) Mixing multi-modal features performs best.}
We observe the single modality, i.e., semantic or relational features, underperforms their combination patterns, i.e., SND-all and RND-all, indicating that the semantic and relational features are complementary to one another.
However, for the RND task, the ad-hoc semantic features alone can compete with their combinations.
The relational features make marginal improvements. That explains why the best contest approach in this task doesn't take advantage of relational features. Therefore, 
how to effectively incorporate relational aspects is still an open question. 

\begin{figure}[t]
	\centering
	\subfigure[SND.]{\label{subfig:sna_cases}
		\includegraphics[width=0.23\textwidth]{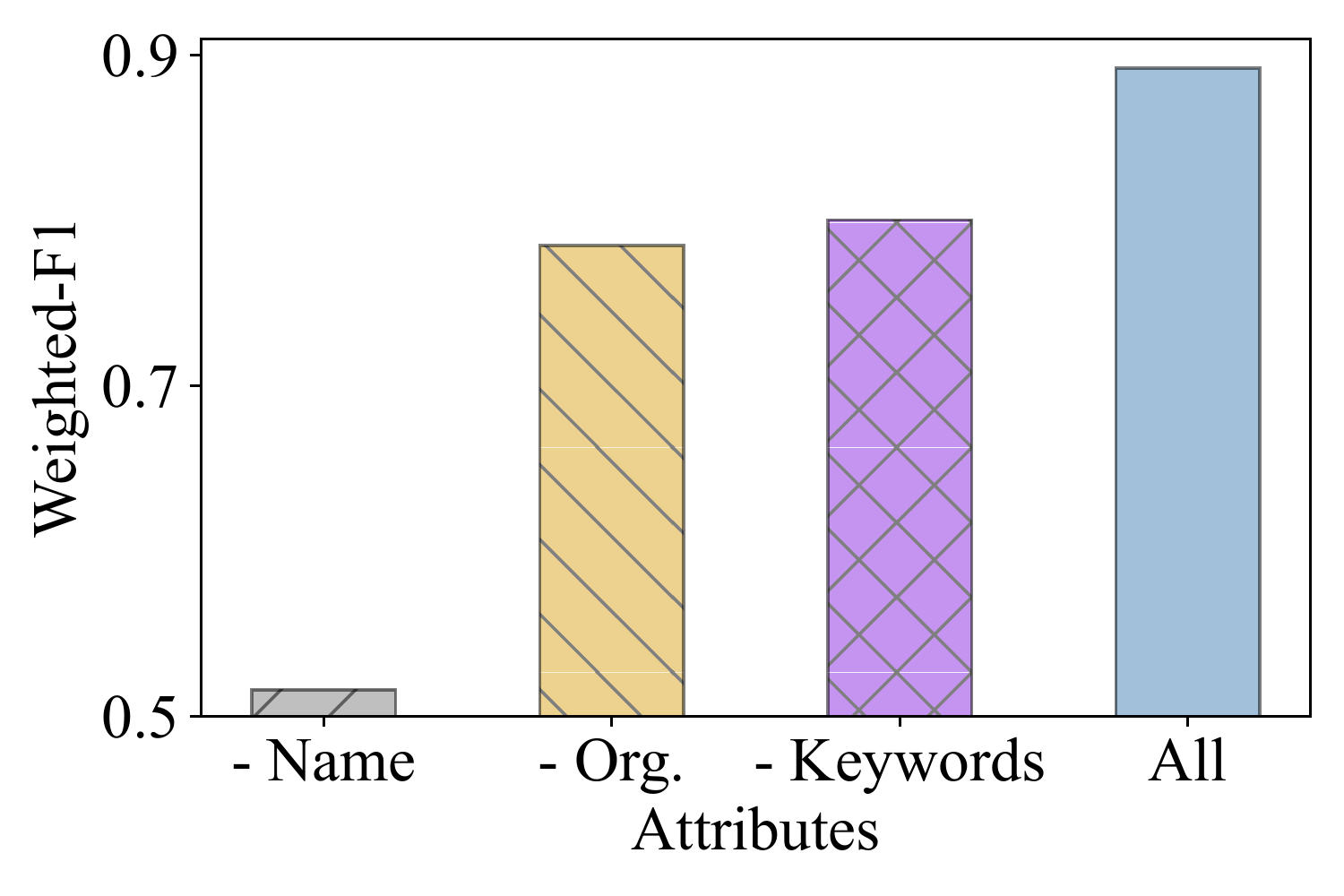}
	}
	\hspace{-0.1in}
	\subfigure[RND.]{\label{subfig:rna_cases}
		\includegraphics[width=0.23\textwidth]{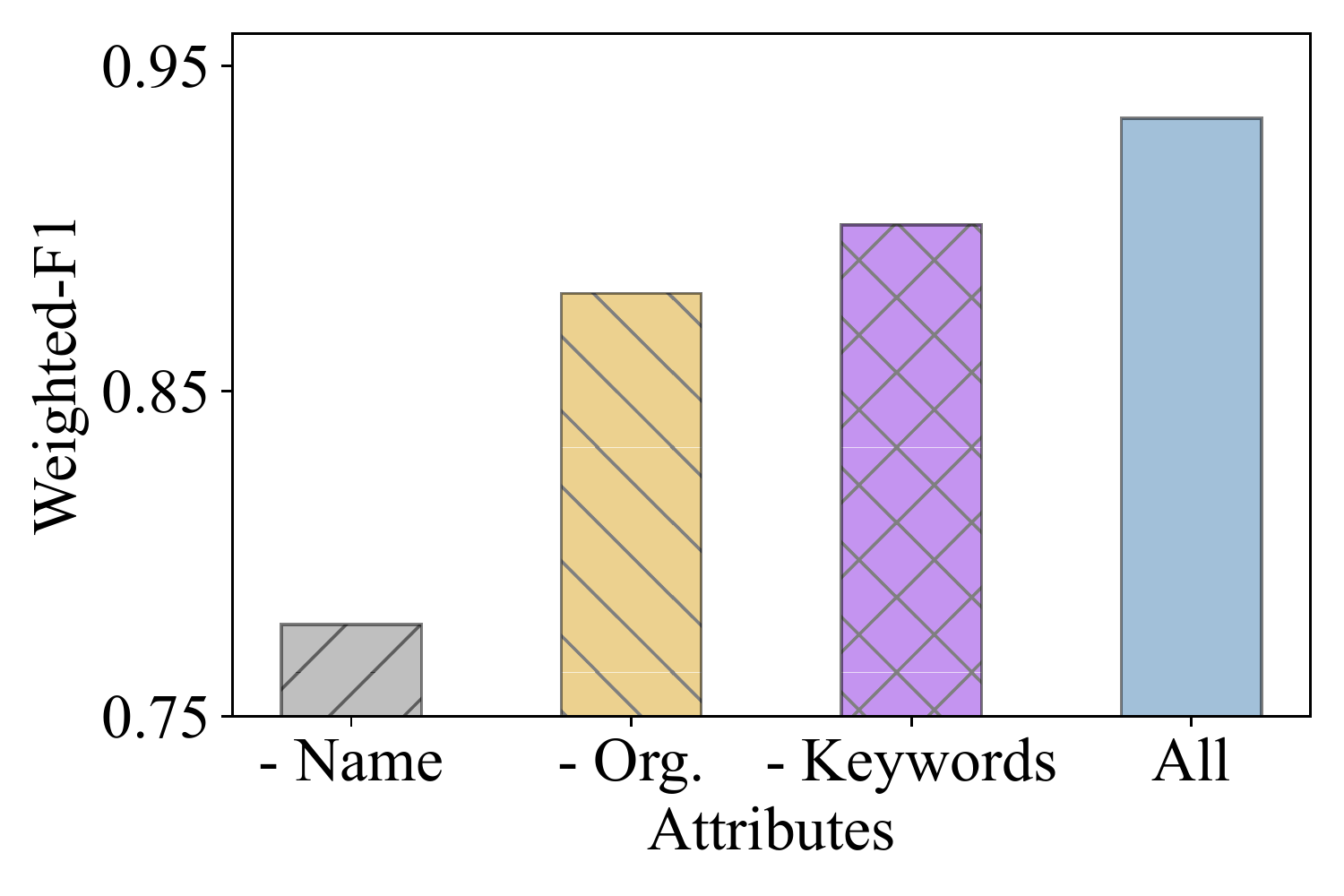}
	}
	\vspace{-10pt}
	\caption{\label{fig:hard_cases} \textbf{Realistic cases analysis.}}
	\vspace{-10pt}
\end{figure}


\subsection{Overall Evaluation}
In this section, we compare the proposed SND-all and RND-all frameworks with existing state-of-the-art name disambiguation methods of the SND and RND tasks\footnote{We only consider the baselines with the released code.}, respectively. The experimental results are performed on the \da-v3 test set. 

\vpara{Compared Baselines.}
Besides the methods of the contest winner, we also compare other prevailing methods,

\ipara{From-scratch Name Disambiguation.} 
\textbf{G/L-Emb}~\cite{zhang2018name} learns paper embeddings on a global paper-paper network and then fine-tunes the embeddings on a local paper-paper network built for each name by graph auto-encoding. 
\textbf{LAND}~\cite{santini2022knowledge} constructs the heterogeneous knowledge graphs (KGs) with papers and authors and leverages KGs embedding techniques to obtain the node embeddings, based on which it performs clustering methods.
\textbf{IUAD}~\cite{li2021disambiguating} determines the authorship of papers via reconstructing the collaboration network where nodes are authors and edges are the coauthor relationships.
\textbf{SND-all} is our proposed strong baseline based on the empirical studies in \secref{subsec:mix_feat}. It mixes soft semantic features with heterogeneous relational graph features to perform the SND task.

\ipara{Real-time Name Disambiguation.} We adopt the following baselines,
\textbf{IUAD}~\cite{li2021disambiguating} is also employed to perform the RND task via reconstructing the collaboration network between newly-arrived papers and existing authors.
\textbf{CONNA}~\cite{chen2020conna}
is an interaction-based model. The basic interactions are built between the token embeddings of two attributes, then different attributes matrices are aggregated as the paper-level interactions, and finally, the paper-level matrices are aggregated as author-level interactions, and \textbf{CONNA+Ad-hoc.} is also a combination methodology that incorporates hand-crafted features into CONNA framework introduced in~\cite{chen2020conna}. For fair comparisons, we leverage features used in Table~\ref{tb:handfeatures}.
\textbf{RND-all} is also our proposed method based on the findings in \secref{subsec:mix_feat}. It adopts the soft and ad-hoc semantic features used in \secref{subsubsec:semantic}. It also builds heterogeneous ego-graphs as relational features. The two features are combined to make predictions.

Other prevailing methods, such as Louppe et al.~\cite{louppe2016ethnicity}, Zhang et al.~\cite{zhang2017name}, Camel~\cite{zhang2018camel}, etc, are empirically proven to be less powerful than the adopted baselines, and thus are ignored in the experiments.

\begin{table}
	\newcolumntype{?}{!{\vrule width 1pt}}
	\newcolumntype{C}{>{\centering\arraybackslash}p{4em}}
	\caption{
		\label{tb:overall_sna} Performance of from-scratch name disambiguation (\%).
	}
	\footnotesize
	\centering 
	\renewcommand\arraystretch{1.0}
	\begin{tabular}{@{~}l?@{~}*{1}{CCC}@{~}}
		\toprule
		\textbf{Model} &\textbf{Pairwise-Precision} &\textbf{Pairwise-Recall} &\textbf{Pairwise-F1}
		\\
		\midrule

		G/L-Emb 
		& 50.77 & 84.64 & 63.48\\
        LAND 
        &61.20 & 61.12 & 61.12
        \\
        IUAD
        &58.82 & 65.22 & 61.63
        \\			
		Contest Winner
		& 82.72	& \textbf{96.59}	& 89.14\\
		\midrule
		SND-all  
		& \textbf{83.06}	&96.35	& \textbf{89.22}  \\

		\bottomrule
	\end{tabular}
	
\end{table}

\begin{table}
	\newcolumntype{?}{!{\vrule width 1pt}}
	\newcolumntype{C}{>{\centering\arraybackslash}p{4em}}
	\caption{
		\label{tb:overall_rna} Performance of real-time name disambiguation (\%).
	}
	\footnotesize
	\centering 
	\renewcommand\arraystretch{1.0}
	\begin{tabular}{@{~}l?@{~}*{1}{CCC}@{~}}
		\toprule
		\textbf{Model} & \textbf{Weighted-Precision} & \textbf{Weighted-Recall} & \textbf{Weighted-F1} \\
		\midrule
		
		
		IUAD
		& 75.53 & 90.49 & 82.34
		\\
        
		CONNA
		& 90.54 & 89.22 & 89.64
		\\
		CONNA+Ad-hoc.
		& 90.23 & 92.64 & 91.14
		\\
		Contest Winner
		&92.09	& \textbf{94.95}	& 93.49\\
		\midrule
		RND-all  
		& \textbf{92.14} & 94.94	& \textbf{93.52}  \\
		
		\bottomrule
	\end{tabular}
\end{table}

\vpara{Results.} Table~\ref{tb:overall_sna} and Table~\ref{tb:overall_rna} demonstrate the performance of various name disambiguation methods on the two tasks. 
The proposed SND-all, RND-all, and the contest winner significantly outperform other baselines by 25.74$\sim$28.10\%  pairwise-F1 and 2.35$\sim$11.80\% weighted-F1 respectively.
The significant performance gap between our proposed method and baselines proposed in recent research sheds light on the capability of prevailing name disambiguation methods is still far from satisfactory, which also reflects the significance of the \sda benchmark.
Moreover, our proposed simple yet effective methods slightly outperform the contest winner method, suggesting that our empirical factor analysis successfully captures the essential components that enhance the effectiveness of name disambiguation methods.    

\subsection{Performance in Realistic Cases}
Papers in the \sda benchmark always contain rich information since annotators prefer to work on papers owning abundant attributes that provide helpful evidence to support their decisions.
Unfortunately, online digital libraries always contain a lot of papers with sparse attributes, meaning that papers with multiple attributes are absent. 
Taking AMiner for example, almost half of the newly-arrived papers lack the attributes of organizations. 
To understand the online name disambiguation scenarios on these papers, we perform SND-all and RND-all on these sparse-attributes cases.

\vpara{Results.} The results are shown in \figref{fig:hard_cases}. 
Among these, the papers without author names perform worst, dropping 36.72\% pairwise-F1 and 15.58\% weighted-F1. 
The absence of the attribute of organizations or keywords also significantly degenerates the online performance of name disambiguation algorithms on both tasks by dropping 8.97$\sim$10.51\% pairwise-F1 and 4.28$\sim$6.40\% weighted-F1. The results indicate that the online name disambiguation scenario is even more sophisticated than what we show on \da. We will update datasets with sparse attributes to encourage more real-world online name disambiguation scenarios in the future.

		
			
		

\begin{figure}[t]
	\centering
	\includegraphics[width=0.48\textwidth]{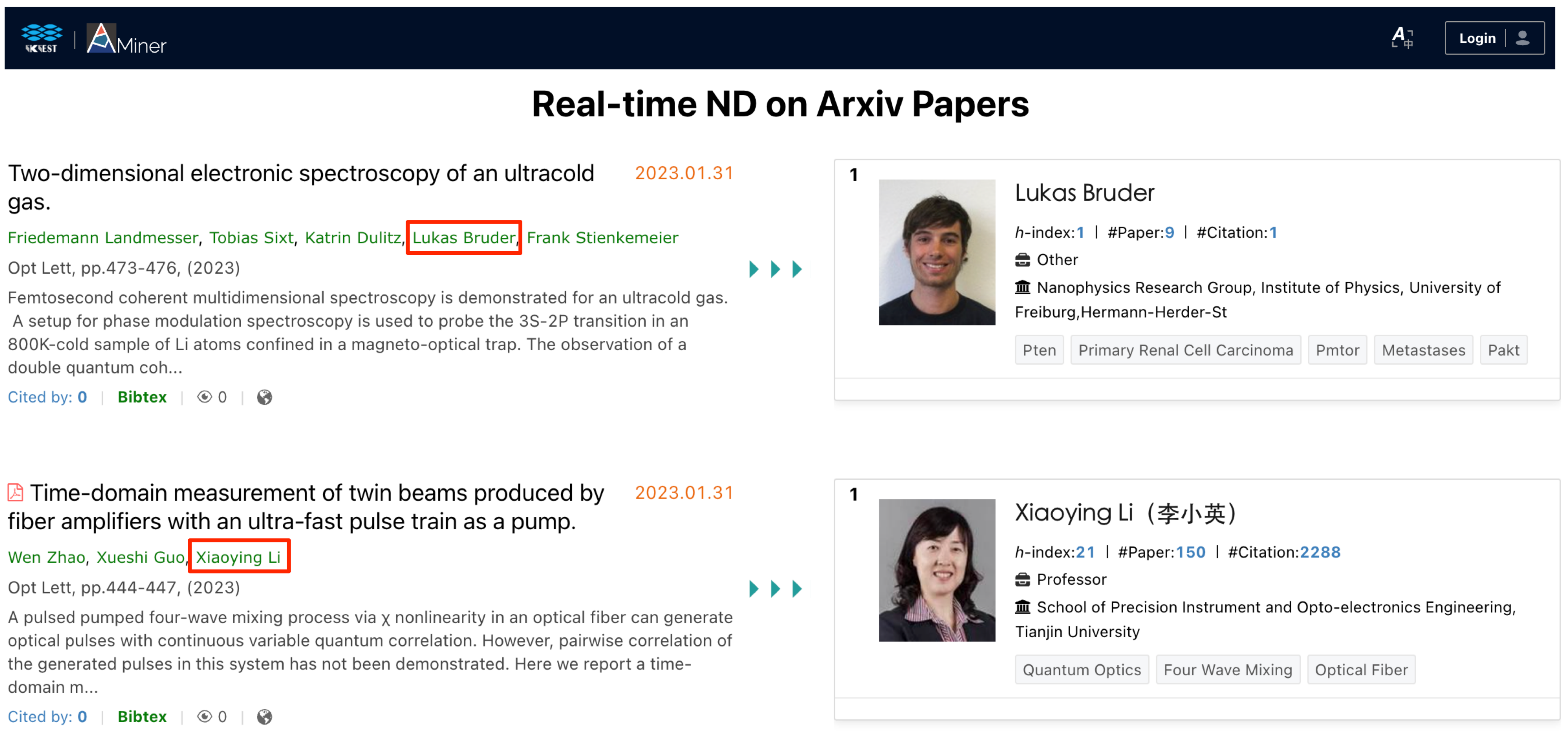}
	\vspace{-20pt}
	\caption{\label{fig:arxiv_demo} A demo about  disambiguating daily papers from arXiv.org.}
	\vspace{-10pt}
\end{figure}


\section{\sda toolkit}
\begin{figure*}[t]
	\centering
	\includegraphics[width=0.85\textwidth]{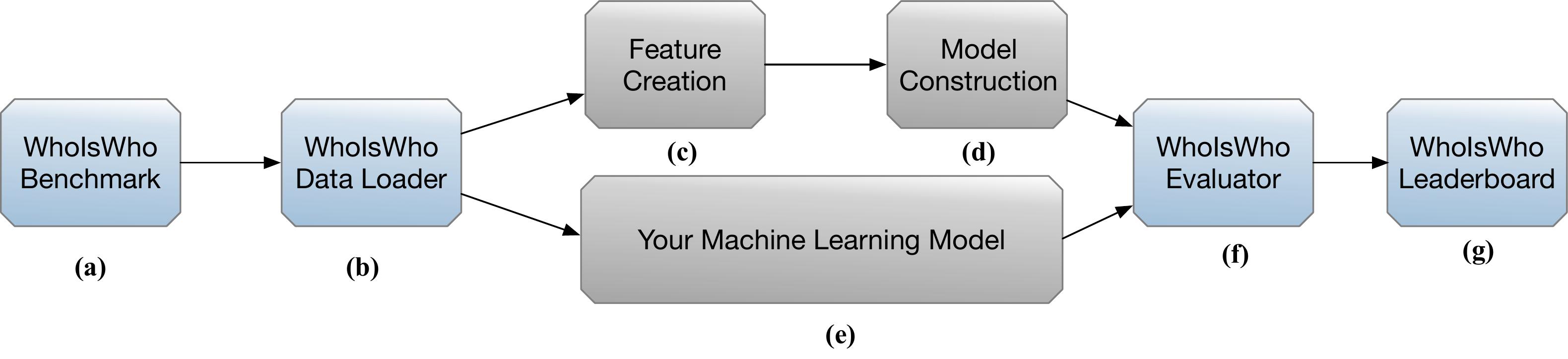}
	\vspace{-10pt}
	\caption{\label{fig:toolkit} Overview of the \sda toolkit pipeline.
	\textmd{\textbf{(a)} \sda provides the large-scale benchmark with high ambiguity and large quantity. \textbf{(b)} The \sda toolkit automates dataset processing and splitting. That is, the data loader automatically loads arbitrary versions of datasets, and further split the datasets in a standardized manner. \textbf{(c)} \sda toolkit provides flexible modules for feature creation including semantic features characterization and relational graph construction, based on that  \textbf{(d)} researchers can adopt models pre-defined in the toolkit library for training and prediction.  Moreover, \textbf{(e)} researchers can build their own feature processing process and develop ML models.  \textbf{(f)} \sda evaluates the model in a task-dependent manner and outputs the model performance on the validation set. 
	Finally,  \textbf{(g)} \sda provides public leaderboards to keep track of recent advances.
	}}
	\vspace{-10pt}
\end{figure*}

By automating 
data loading, feature creation, model construction, and evaluation processes,
the \sda toolkit 
is easy for researchers to use and let them develop new name disambiguation approaches.
The overview of the toolkit pipeline is illustrated in \figref{fig:toolkit}.
The toolkit is fully compatible with PyTorch and its associated deep learning libraries, such as Hugging face~\cite{wolf-etal-2020-transformers}. 
Additionally, the toolkit offers library-agnostic dataset objects that can be used by any other Python deep learning frameworks such as Tensorflow~\cite{abadi2016tensorflow}.  
To keep things simple, we concentrate on building a basic RND method using PyTorch shown in Listing~\ref{lst:imple_codes}. More details refer to \url{https://github.com/THUDM/WhoIsWho}.

\vpara{Disambiguating Arxiv Papers.} 
We deploy the RND-all method implemented by our toolkit on AMiner to disambiguate daily papers from arXiv.org on-the-fly. A demo page is depicted in \figref{fig:arxiv_demo}.  The details refer to \secref{sec:arxiv_deploy}.
We manually check the latest 100 disambiguation results reflecting that 90\% assignments are accurate.



\section{Related Work}
\label{sec:related}
Here, we recall the prevailing name disambiguation datasets and the state-of-the-art name disambiguation algorithms. 

\vpara{Name Disambiguation Datasets.} 
The size of datasets heavily influences the performance of name disambiguation algorithms. 
To address the problem, 
the community has created a large number of name disambiguation datasets recently. Among them, several efforts directly harvest datasets from existing digital libraries, including PubMed~\cite{zhang2020mining, zeng2020large}, DBLP~\cite{kim2018evaluating}, etc.~\cite{zhang2021lagos, wang2018acekg, kim2021orcid}. 
However, the assignment mistakes, as shown in \figref{fig:whoiswho_demo}, hamper the development of effective algorithms~\cite{chen2022gccad, zhang2021name}. Others attempt to manually label a small amount of data based on noisy data from existing databases to reduce data noises~\cite{song2015exploring, vishnyakova2019new, han2005name, qian2015dynamic,tang2012unified, wang2011adana, muller2017data, kang2011construction, zhang2018name, louppe2016ethnicity}.
Most of them, however, do not have sufficient instances, as shown in \figref{fig:data}.  The detailed data statistics refer to Table~\ref{tb:overall_sta}. Some of them have restricted scopes, for example, SCAD-zbMATH~\cite{muller2017data} is customized for a mathematical domain.
The fragile inductive bias affects the performance and generalization of name disambiguation methods that are trained on these datasets. 
\citet{subramanian2021s2and} build a unified dataset via aggregating several small scales of datasets.
However, the quality of constituents has not been checked.

\vpara{Practical Tasks \& Algorithms.}
Most efforts focus on the SND task.
Generally, they operate via three steps:
blocking, paper similarity matching, and clustering.
\citet{backes2018impact} discusses the name-blocking step.
Several works lay emphasis on paper similarity matching and clustering steps.
Early attempts designed hand-crafted similarity metrics~\cite{cota2010unsupervised} to measure paper similarities.
Then, researchers discover that constructing paper similarity graphs excels at learning high-order similarity~\cite{fan2011graph,dong2017metapath2vec,qiao2019unsupervised,kang2011construction,zhang2018name}. 
As for clustering steps, 
the clustering methods such as hierarchical agglomerative clustering and DBSCAN are adopted.
Among them, DBSCAN is preferred by practitioners as there is no need to specify the cluster number.

The RND task, which aims to assign newly-arrived papers to existing authors, is a more practicable scenario for online academic systems,
Besides adopted baselines,
\citet{qian2015dynamic} predict the likelihood of a paper being written by a specific author via the attributes of coauthor and keyword.
\citet{pooja2022online} utilize dynamic graph embedding to model evolving graphs.
Several works~\cite{li2021disambiguating,zhang2019dirichlet} further employ a probabilistic model for online paper assignments.

Inevitable cumulative errors  will greatly affect the efficacy of name disambiguation algorithms. Thus, the IND task is vital to guarantee the reliability of academic systems. Unfortunately, the issue has not received much attention~\cite{chen2022gccad}.  

Previous methods are usually evaluated on diverse small-scale datasets, which hamper the development of the community. Thus, a large-scale benchmark, a regular leaderboard with comprehensive tasks, together with an easy-to-use toolkit for web-scale academic name disambiguation should be concerned.

\section{Conclusions}
\label{sec:con} 

This paper delivers \sda including a benchmark, a leaderboard, and a toolkit for web-scale academic name disambiguation.
Specifically, the large-scale benchmark with high ambiguity enables the devising of robust algorithms. Sponsored contests with two tracks promote the advances of the name disambiguation community. A regular leaderboard is publicly available to keep track of recent advances.  An easy-to-use toolkit is designed to allow end users to rapidly build their own algorithm and publish their results on a regular leaderboard that records recent advances. In summary, \sda is an ongoing, community-driven, open-source project. 
We also encourage contributions from the community.

\small
\vpara{Acknowledgments.} 
This work was supported by Technology and Innovation Major Project of the Ministry of Science and Technology of China under Grant 2020AAA0108400 and 2020AAA0108402, 
the NSF of China for Distinguished Young Scholars (No. 61825602), NSF of China (No. 62076245, 62276148), CCF-Zhipu202306, and the Public Computing Cloud at Renmin University of China.
\normalsize

\clearpage
\bibliographystyle{ACM-Reference-Format}
\balance
\bibliography{reference} 
\clearpage
\appendix

\begin{lstlisting}[language=Python,breaklines=True,frame=single,numbers=none,caption=Basic RND algorithm,label={lst:imple_codes}]
# Module-1: Data Loading
from whoiswho.dataset import LoadData, SplitDataRND
# Load specific versions of dataset.
train = LoadData(name="v3", type="train", partition=None)
# Split data into unassigned papers and candidate authors
unassigns, candidates = SplitDataRND(train, split="time", ratio=0.2)

# Modules-2: Feature Creation
from whoiswho.featureGenerator import AdHocFeatures
# Extract default n-dimensional ad-hoc features.
pos_feats, neg_feats = AdHocFeatures(unassigns, candidates, feature_mode="default", negatives=3)

# Module-3: Model Construction
from whoiswho.loadmodel import ClassficationModels
# build a basic classfication model.
predictor=ClassficationModels(type="MLP", ensemble=False)
# Automatic training 
from whoiswho.training import AutoTrainRND
predictor = AutoTrainRND(inputs = (pos_feats, neg_feats), predictor, epoch=1, bs=1, early_stop=None)

# Modules-4: Evaluation on the validation data
# Load validation data
unassigns, candidates, gt = LoadData("v3", type="Valid", task="RND")
# Assign unassigned papers
assign_res = predictor.predict(unassigns, candidates)
# Evaluate the RND results
from whoiswho.evaluation import RNDeval
weighted_Precision, weighted_Recall, weighted_F1 = RNDeval(assign_res, gt)
\end{lstlisting}

\section{Appendices}
\subsection{WhoIsWho Toolkit Pipeline.}
\figref{fig:toolkit} demonstrates the overview of the \sda toolkit pipeline. A toy example of building basic RND algorithms is shown in Listing~\ref{lst:imple_codes}.

\subsection{Data Organizations in \sda Benchmark.}
To date, \sda released three versions of datasets, i.e., \da-v1, -v2, and -v3, with one specified dataset, \da-v3.1, for the IND task. Among them, v1,v2,v3 datasets have the same organizations with the SND and RND tasks. Here, we briefly review the data organizations.
\subsubsection{\da-v1/v2/v3} 
The datasets are organized into the format of a two-level dictionary, i.e., names-authors-papers as shown in Listing~ \ref{lst:whoiswho_v1}. The key of the first-level dictionary is author names and the value is author profiles with the ``same name''. The term ``same name'' refers to the ways to unify names using name blocking techniques~\cite{backes2018impact,kim2018evaluating},
such as moving the last name to the first or preserving all name initials but the last name. For example, the variants of ``Jing Zhang'' are ``Zhang Jing'', ``J. Zhang'', and ``Z. Jing''. The author profiles are also organized as a dictionary with the key being author IDs and the value being paper IDs of the author.
For each paper, we collect the title, author names, organizations of all the authors, keywords, abstract, publication year, and venue (conference or journal) as its attributes. A toy example of the paper with ID ``9PgiwDo7'' is shown in Listing~\ref{lst:whoiswho_paper}.

\subsubsection{\da-v3.1 for the IND task}
\da-v3.1 is organized as a one-level dictionary, as the IND task aims to detect and remove the error papers within each author. The key of the dictionary is author IDs, and the value is the papers belonging to the author, i.e., the normal data, and the manually detected error papers, i.e., the outliers. A demo case is present in Listing~\ref{lst:whoiswho_v3.1}.

\subsection{Data Split of \sda Contest}
We claim the process of splitting the \sda datasets into training, validation, and test sets for the contest of three tasks, i.e., SND, RND, and IND, respectively.

\ipara{From-scratch Name Disambiguation.} The SND task targets at partitioning papers of the author's name into different groups. Each group contains papers from the same author while papers in different groups belong to different authors. Thus, we first split the datasets into training, validation, and test set via the level of author names following specific ratios. Then for the validation and test sets, we delete the authorships between authors and papers in each name as shown in Listing~\ref{lst:whoiswho_snd}. Researchers should correctly cluster papers belonging to the same author into the same group.

\ipara{Real-time Name Disambiguation.} The RND task aims at assigning newly-arrived papers to existing authors. Thus, firstly, we also split the datasets into training, validation, and test sets via the level of author names following specific ratios. Then for the validation and test sets, we sort papers within each author via the published year in ascending order. To simulate the real RND scenario, we treat the latest papers as the new-arrived unassigned papers and the remains as existing author profiles in each author, as shown in Listing~\ref{lst:whoiswho_rnd}. We also add several NIL papers, i.e., papers that can not be assigned to any existing author profiles, to the unassigned papers. Researchers need not only correctly assign papers to the right author, but also to distinguish NIL papers.

\ipara{Incorrect Assignment Detection.} 
The IND task is designed for detecting and removing the error papers within each author.
Concretely, we construct the dataset of the IND task as follows:  
1) as illustrated in Table 1, the overall data annotation pipeline includes a 'Clean' step, during which annotators remove or split obviously incorrect papers from the concerned author. The subsequent 'Validate' step allows annotators to perform the same 'Clean' function on incorrectly assigned papers that are more difficult to identify. These two stages provide a sufficient number of incorrectly assigned papers to be detected in the IND task. 
2) Some authors manually maintain their profiles, such as adding new papers or removing papers that do not belong to them. We also collect the removed papers as targets for detection in the IND task. 
Thereby, we split the training, validation, and test sets via the author groups.

\begin{minipage}{.95\linewidth}
\begin{lstlisting}[language=Python, breaklines=True,frame=single,numbers=none,caption=Data organizations of WhoIsWho-v1/v2/v3,label={lst:whoiswho_v1}]
{
    "guanhua_du": { # Author name,
        "zsOOUZxZ": [ # Author IDs,
            "QDMcmF8V", # Paper IDs,
            "9PgiwDo7"
        ]
    },
    "bin_yu": {
        "HoH18DsE": [
            "VMYs96sn",
            "YT4XzThC",
            "S3RARClD",
            "wM8dXlKT"
        ],
        "WYZVZfO0": [
            "OTvYjfnt",
            "EzzruFin"
        ],
        "9BMlVP0u": [
            "brlzUqnH",
            "HWAfXDPx",
            "RZPhHOMm",
            "vebukM2n"
        ]
    }
}
\end{lstlisting}
\end{minipage}

\begin{minipage}{.95\linewidth}
\begin{lstlisting}[language=Python, breaklines=True,frame=single,numbers=none,caption=Data organizations of WhoIsWho-v3.1,label={lst:whoiswho_v3.1}]
{
    "HoH18DsE": { # Author IDs,
        "name": "xxx", # Name of the author,
        "normal_data":[ # Papers belong to the author,
        "VMYs96sn",
        "YT4XzThC",
        "S3RARClD",
        "wM8dXlKT"
        ],
        "outliers":[ # Papers wrongly assigned to the author, 
        "OTvYjfnt",
        "EzzruFin"
        ]
    },
}
\end{lstlisting}
\end{minipage}

\begin{minipage}{.95\linewidth}
\begin{lstlisting}[language=Python, breaklines=True,frame=single,numbers=none,caption=Data organizations of WhoIsWho-v1/v2/v3,label={lst:whoiswho_paper}]
{
    "9PgiwDo7": {
        "id": "9PgiwDo7",
        "title": "Constrained phase transformation of prestrained TiNi fibers embedded in metal matrix smart composite",
        "abstract": "The reverse martensitic transformation of TiNi fibers embedded in a metal matrix smart composite has been studied. Results show that under the influence of temperature and recovery stress, the reverse martensitic transformation of TiNi fibers can be divided into two parts with different kinetic characteristics: the reverse transformation of self-accommodating martensite and that of oriented martensite. The relationship between martensitic fraction and temperature was calculated.",
        "keywords": [
            "null"
        ],
        "authors": [
            {
                "name": "Yanjun Zheng",
                "org": "Dalian University of Technology(Dalian University of Technology,Dalian Univ. of Technol.),Dalian,China"
            },
            {
                "name": "Lishan Cui",
                "org": "China University of Petroleum - Beijing(China University of Petroleum,University of Petroleum),Beijing,China"
            },
            {
                "name": "Dan Zhu",
                "org": "China University of Petroleum - Beijing(China University of Petroleum,University of Petroleum),Beijing,China"
            },
            {
                "name": "Dazhi Yang",
                "org": "Dalian University of Technology(Dalian University of Technology,Dalian Univ. of Technol.),Dalian,China"
            }
        ],
        "venue": "Materials Letters",
        "year": 2000
    }
}
\end{lstlisting}
\end{minipage}

\begin{minipage}{.95\linewidth}
\begin{lstlisting}[language=Python, breaklines=True,frame=single,numbers=none,caption=Data organizations of the validation/test set in the RND task,label={lst:whoiswho_snd}]
{
    "bin_yu": {
        [
            "VMYs96sn",
            "YT4XzThC",
            "S3RARClD",
            "wM8dXlKT"
            "OTvYjfnt",
            "EzzruFin",
            "brlzUqnH",
            "HWAfXDPx",
            "RZPhHOMm",
            "vebukM2n"
        ]
    }
}
\end{lstlisting}
\end{minipage}

\begin{minipage}{.95\linewidth}
\begin{lstlisting}[language=Python, breaklines=True,frame=single,numbers=none,caption=Data organizations of WhoIsWho-v1/v2/v3,label={lst:whoiswho_rnd}]
# Existing author profiles.
{
    "HoH18DsE": [
        "VMYs96sn",
        "YT4XzThC",
        "S3RARClD",
        "wM8dXlKT"
    ],
    "9BMlVP0u": [
        "brlzUqnH",
        "HWAfXDPx",
        "RZPhHOMm",
    ]
}

# Newly-arrived unassigned papers.
{
    [
    "vebukM2n-1", # "-1" means the 2nd (Author index begins at 0) author should be disambiguated.
    "RZPhHOMm-3"
    ]
}
\end{lstlisting}
\end{minipage}

\subsection{Running Environment}
We implement all the experiments model by PyTorch and run the code on an Enterprise Linux Server with 40 Intel(R) Xeon(R) CPU cores (E5-2640 v4 @ 2.40GHz  and 252G memory) and 1 NVIDIA Tesla V100 GPU core (32G memory).

\subsection{Implementation Details of the SND-all}
\label{sec:snd_details}
Here, we introduce the technical details of the SND-all method, consists of the soft semantic features, the heterogeneous relational graph features, and the combination patterns.
\subsubsection{Soft Semantic features.}
We extract all paper attributes from the \sda benchmark, including title, abstract, venue, keywords, year, and author’s organization, to train the word embedding model using word2vec implemented by genism with embedding dimensions set to 100. 
The hyper-parameters are defined as follows: min\_count=2, window=5, negative=5. 
Then, for each paper, we average the word embedding of the title, keywords, and the organization of the author being clarified as the semantic feature as the final paper embeddings. Specifically, the fields of title and keywords are directly picked from the paper attributes. While for the organization, we extract the organizations of the target author to be disambiguated.

\begin{figure}[t]
	\centering
	\includegraphics[width=0.48\textwidth]{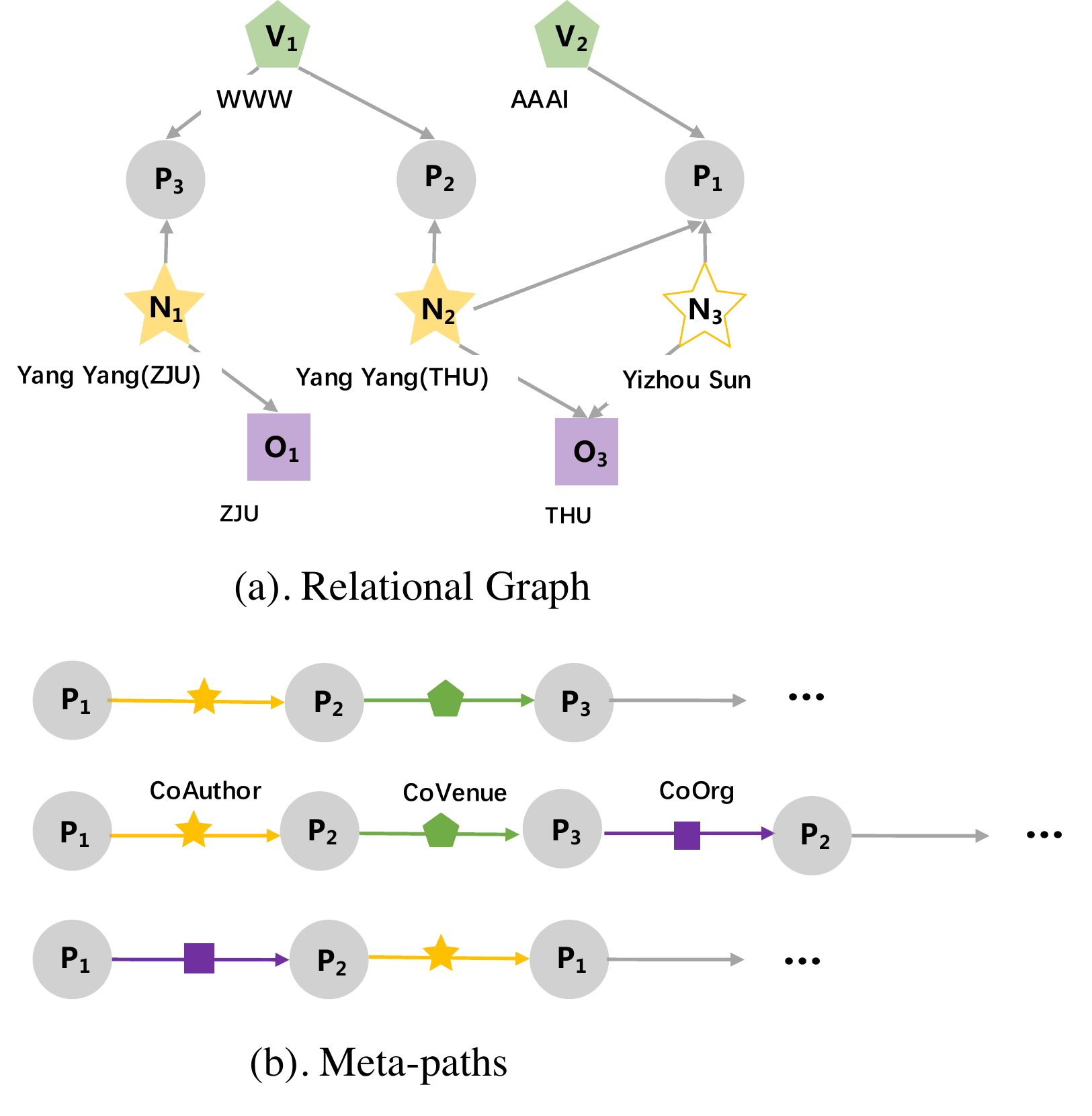}
	\caption{\label{fig:snd_all_relation} Heterogeneous relational graph and meta-paths adopted in the SND-all method.}
\end{figure}

\subsubsection{Relational Semantic Features.}
For each target name, we build a heterogeneous network among papers with three relation types CoAuthor, CoOrg, and CoVenue. The adopted meta-paths are shown in \figref{fig:snd_all_relation}. The weights of the heterogeneous edge are determined by the number of co-occurrences, i.e., the number of co-authors for CoAuthor,  the number of co-occurring words in organization for CoOrg, and also the number of co-occurring words in venue/journal for CoVenue. 
We set a random probability as 0.1 that the paper walks to neighboring papers via the CoVenue edge to reduce the impact of common word noises in the venue.
The walk length is set to 20, and the number of random walks for each node is set to 5. 
Then, we save the collection of paths for each paper and train them using word2vec. 
The isolated papers that lack neighbors are preserved for post-match processing.

\subsubsection{Feature Combinations.} 
Finally, we calculate pairwise distances among papers via the semantic and relational embeddings of papers, respectively.
Then we simply add the two similarity matrices to form the combined similarity matrix. 
To obtain the clustering results, we adopt DBSCAN to generate clusters with db\_eps as 0.2 and db\_min as 4. 
For outliers that can not be assigned to any groups, 
we conduct a rule-based method for post-matching. Specifically, we conduct character matching on names of co-authors as CoA, titles, and keywords of two papers as CoW, and we adopt tanimoto distance for calculating the similarity of organizations and venues as CoO and CoV. CoA, CoW, CoO, and CoV are added with the weights of \{1.5, 0.33, 1.0, 1.0\} respectively. 
Then we assign the papers to the group if the score is above a pre-defined threshold, i.e., 1.5 in our method. 

\begin{table}[t]
		{\caption{\label{tb:handfeatures}}The detailed definitions of 36-dimensional hand-crafted features. \small{$p$: target paper, $a$: target author in $p$, $c$: candidate person.} }
		\vspace{-0.08in}
		{
		\small
		\renewcommand{\arraystretch}{1}%
			{
				\setlength{\extrarowheight}{1pt}
				\begin{tabular}{
						@{}c@{ } l@{}}
					\noalign{ \hrule height 1pt}
					\textbf{No.}   & \textbf{Feature description} \\ \hline
					\textbf{1}      &  TF-IDF score of $a$'s coauthors in $c$\\ 
                    \textbf{2}      &  TF-IDF score of $a$'s coauthors in $c$ multiplied by co-occurrence \\ & times in $c$ \\ 
                    \textbf{3}      &  Ratio of $a$'s coauthors in $p$'s author names\\    
                    \textbf{4}      &  Ratio of $a$'s coauthors  in $c$'s author names\\ \hdashline 

                    \textbf{5}      &  TF-IDF score of $a$'s title common part in $c$\\ 
                    \textbf{6}      &  TF-IDF score of $a$'s title common part in $c$ multiplied by \\ & co-occurrence times in $c$ \\ 
                    \textbf{7}     &  Ratio of $a$'s title common part in $a$'s title \\ 
                    \textbf{8}      &  Ratio of $a$'s title common part  in $c$'s titles\\  

					\textbf{9}      &  Max Jaccard similarity  between $a$'s title and $c$'s titles\\ 
                    \textbf{10}      &  Mean Jaccard similarity  between $a$'s title and $c$'s titles\\  
                    \textbf{11}      &  Max Jaro–Winkler similarity  between $a$'s title and $c$'s titles\\  
                    \textbf{12}      &  Mean Jaro–Winkler similarity  between $a$'s title and $c$'s titles\\  	\hdashline 			

                     \textbf{13}      &  TF-IDF score of $a$'s venue common part in $c$\\ 
                    \textbf{14}      &  TF-IDF score of $a$'s venue common part in $c$ multiplied by \\ & co-occurrence times in $c$ \\ 
                    \textbf{15}     &  Ratio of $a$'s venue common part in $a$'s venue \\ 
                    \textbf{16}      &  Ratio of $a$'s venue common part in $c$’s venues \\  

					\textbf{17}      &  Max Jaccard similarity  between $a$'s venue and $c$'s venues\\ 
                    \textbf{18}      &  Mean Jaccard similarity  between $a$'s venue and $c$'s venues\\  
                    \textbf{19}      &  Max Jaro–Winkler similarity  between $a$'s venue and $c$'s venues\\  
                    \textbf{20}      &  Mean Jaro–Winkler similarity  between $a$'s venue and $c$'s venues\\  	\hdashline 

                     \textbf{21}      &  TF-IDF score of $a$'s organization common part in $c$\\ 
                    \textbf{22}      &  TF-IDF score of $a$'s organization common part in $c$ multiplied by \\ & co-occurrence times in  $c$ \\ 
                    \textbf{23}     &  Ratio of $a$'s organization common part in $a$'s organization \\ 
                    \textbf{24}      &  Ratio of $a$'s organization common part in $c$'s  organizations \\  

					\textbf{25}      &  Max Jaccard similarity  between $a$'s organization and $c$'s organizations\\ 
                    \textbf{26}      &  Mean Jaccard similarity  between $a$'s organization and $c$'s organizations\\  
                    \textbf{27}      &  Max Jaro–Winkler similarity  between $a$'s organization and $c$'s organiza \\ & -tions\\  
                    \textbf{28}      &  Mean Jaro–Winkler similarity  between $a$'s organization and $c$'s organiza \\ & -tions\\  	\hdashline 	

                    \textbf{29}      &  TF-IDF score of $a$'s keywords common part in $c$\\ 
                    \textbf{30}      &  TF-IDF score of $a$'s keywords common part in $c$ multiplied by \\ & co-occurrence times in $c$ \\ 
                    \textbf{31}     &  Ratio of $a$'s keywords common part in $a$'s keywords \\ 
                    \textbf{32}      &  Ratio of $a$'s keywords common part in $c$'s keywords\\  

					\textbf{33}      &  Max Jaccard similarity  between $a$'s keywords and $c$'s keywords\\ 
                    \textbf{34}      &  Mean Jaccard similarity  between $a$'s keywords and $c$'s keywords\\  
                    \textbf{35}      &  Max Jaro–Winkler similarity  between $a$'s keywords and $c$'s keywords\\  
                    \textbf{36}      &  Mean Jaro–Winkler similarity  between $a$'s keywords and $c$'s keywords\\  	
                     			
					\noalign{\hrule height 1pt}
				\end{tabular}}
				
			}
\end{table}

\subsection{Implementation Details of RND-all}
\label{sec:rnd_details}
Here, we introduce the technical details of the RND-all method, which consists of the soft semantic features, the ad-hoc semantic features, the ego graphs, and the combination patterns.

\subsubsection{Soft Semantic Features}
\label{subsec:rnd_all_soft}
Similarly, we obtain the soft semantic embeddings of papers like the SND-all does while leveraging OAGBERT~\cite{liu2022oag}, based on which we get the soft semantic similarities between the target paper and each paper of the candidate author.

\subsubsection{Ad-hoc Semantic Features}
\label{subsec:hands}
Table~\ref{tb:handfeatures} presents the detailed definitions of 36-dimensional hand-crafted features between each unassigned paper and candidate author pair.

\subsubsection{Relational Ego Graph Features}
\label{subsec:rnd_all_ego}
\begin{figure}[t]
	\centering
	\includegraphics[width=0.48\textwidth]{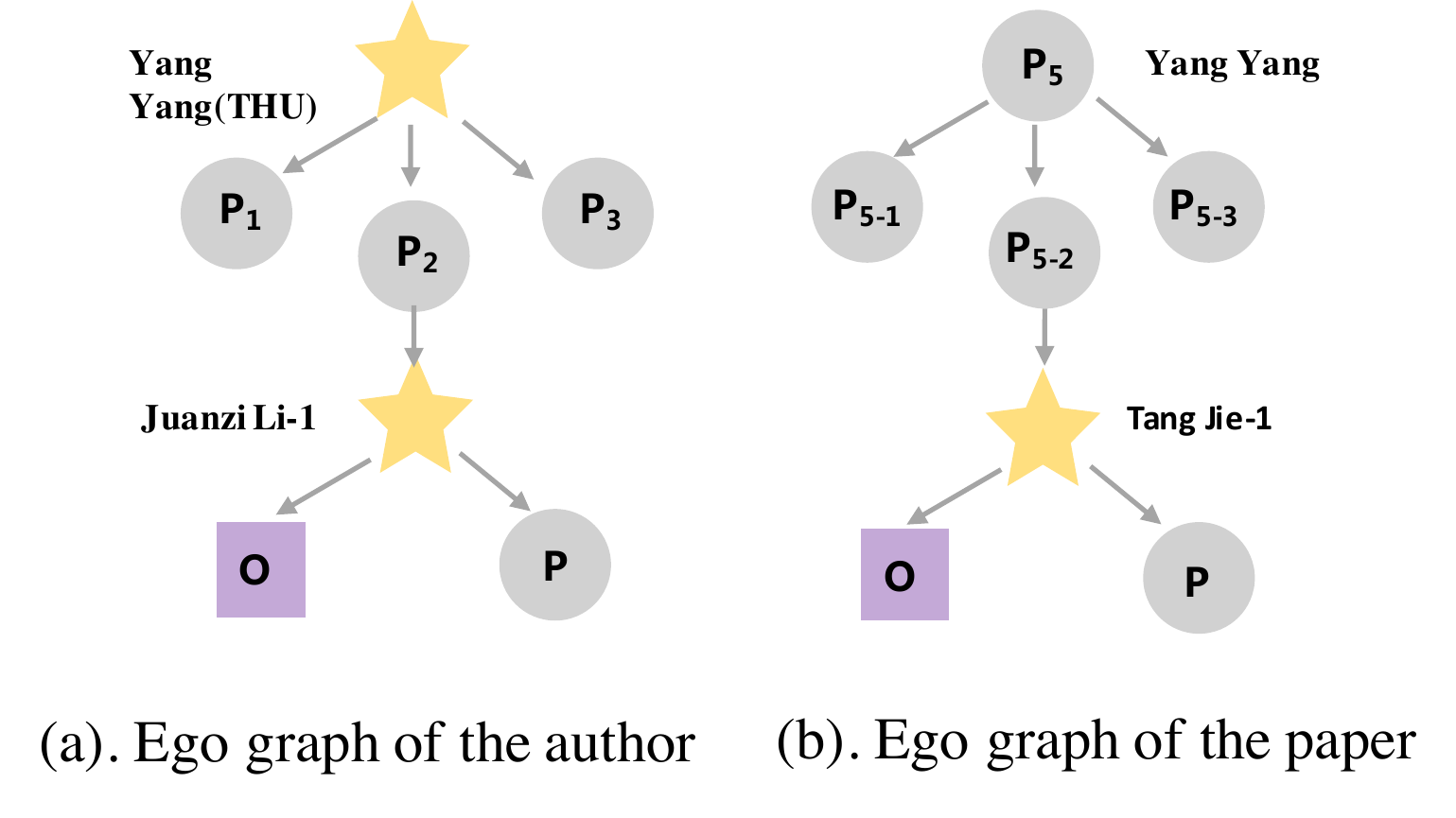}
	\caption{\label{fig:rnd_all_ego} The built ego graph on the target paper and the candidate author in the RND-all method.}
\end{figure}

We collect historical bibliographic data to construct heterogeneous ego-graphs for target papers and candidate authors, with authors, papers, and organizations  as nodes. As shown in \figref{fig:rnd_all_ego}, 
For the concerned center node of the author, we first use OAGBERT to get the paper embedding, then we average all the paper embeddings belong the author as the author embeddings.
After that, we obtain the updated paper or author embeddings via training the graph attention networks (GAT), based on which we get the relational similarities between the center paper and the center authors.



\subsubsection{Feature Combinations}
For the soft semantic features and relational ego graph features, we adopt an RBF kernel function used in \cite{chen2020conna} to extract 41 dimensional aggregated features based on the similarity scores between the target paper and the candidate author.
For the ad-hoc semantic features, we directly adopt 36-dimensional hand-crafted features.
Finally, we simply concatenate aggregated features from the three feature modalities to obtain the 118-dimensional features between the target paper and the candidate author, and adopt ensemble GBDT models to make predictions.

\subsection{Implementation Details of baseline methods}
\label{subsec:baselines}
Here we elaborate on the details of name disambiguation baselines used in the paper. 
\subsubsection{Baselines of the SND task}
\noindent \textbf{G/L-emb}. Is a method for constructing paper-paper graphs using co-authorship connections. The process begins by generating initial paper embeddings through a weighted average of Word2Vec embeddings for all tokens within a paper. Next, the method fine-tunes these embeddings by first learning on a global paper-paper network and then adapting them to a local paper-paper network, specific to each author, using graph auto-encoding. Lastly, a hierarchical agglomerative clustering algorithm (HAC) is employed to segregate these papers into distinct groups.
\noindent \textbf{LAND}. Is a method for constructing heterogeneous scholarly knowledge graphs (KGs) that encompass multiple entities, including papers, authors, venues, affiliations, and more. These KGs also feature various relations, such as co-authorship, and publication venues. Entity input embeddings are initialized using BERT models, and KG embedding techniques are then applied to derive paper and author embeddings. Finally, hierarchical agglomerative clustering (HAC) methods are utilized to group these entities based on the embeddings.
\noindent \textbf{IUAD}. Is a method that constructs collaboration networks by treating papers as nodes and creating edges between two papers if they share the same author. It employs probabilistic generative methods to determine whether two papers in the collaboration network belong to a single author, with the goal of reconstructing the complete collaboration network accurately. Once the probabilistic models are trained, they are utilized to perform SND (Supervised Name Disambiguation) tasks, enhancing the overall disambiguation process. 
\noindent \textbf{SND-all}. Is a proposed baseline method based on empirical studies from contest-winning approaches. It starts by estimating semantic correlations among papers to be disambiguated, using title, keywords, and organizations as soft semantic features. For each paper, SND-all projects these features into corresponding word embeddings via Word2Vec and averages them to create paper embeddings. Cosine similarities are then calculated between papers based on these semantic embeddings.
Additionally, SND-all constructs a heterogeneous network featuring three relational edges among papers: co-author, co-organization, and co-venue. The method employs metapath2vec to generate relational embeddings of papers, and computes relational similarity scores based on these embeddings.
Lastly, SND-all combines both multi-modal similarities to calculate overall similarities among papers. DBSCAN is used to derive the final clustering results based on these similarities.
\noindent \textbf{Contest Winner Method}: Follows a similar pipeline to SND-all. However, it uses all paper attributes to estimate semantic correlations among papers, which has proven to be less effective than utilizing a few informative attributes, such as title, keywords, and organizations, as implemented in the SND-all method. Furthermore, when estimating relational similarities among papers, the Contest Winner Method only considers co-author and co-organization relational edges, offering a more limited perspective compared to SND-all.

\subsubsection{Baselines of the RND task}
\noindent \textbf{IUAD}. Is also employed for the RND task by reconstructing the collaboration network between newly-arrived papers and existing authors.
\noindent \textbf{CONNA}. Is a bottom-up, interaction-based model. To determine similarities between the target paper and candidate authors, it first constructs basic interactions between token embeddings of corresponding attributes for the target paper and each paper of the candidate author. Then, different attribute matrices are aggregated to form paper-level interactions. Specifically, CONNA uses all paper attributes, treating author names as one field and all other attributes as another field. Finally, Learning-to-Rank techniques are used to score the agreements between the target paper and all candidate authors based on the aggregated paper-level matrices.
\noindent \textbf{CONNA+Ad-hoc}. Is a combination methodology that integrates hand-crafted features into the CONNA framework. For fair comparisons, the same ad-hoc features used in RND-all are leveraged.
\noindent \textbf{RND-all}. Is another proposed method based on ablation study findings. It utilizes the soft and ad-hoc semantic features described in Section 4.1. Additionally, to incorporate relational features, RND-all constructs heterogeneous ego-graphs for the target paper and all candidate authors, as shown in Figure 11 of the appendix. A Graph Attention Network (GAT) is employed to obtain fused center node embeddings for the target paper and candidate authors. The relational correlations between the target paper and each candidate author are determined by calculating cosine similarities between the center node embeddings. Lastly, RND-all combines the soft semantic features, ad-hoc semantic features, and proposed relational features to make predictions.
\noindent \textbf{Contest Winner Method}. Estimates the agreement between the target paper and each candidate author using only soft semantic features and ad-hoc semantic features. It also employs all paper attributes when constructing semantic features, which has been proven less effective than adopting a few informative attributes, as used in the RND-all method.

\subsection{Online deployment of disambiguating daily papers from arXiv.}
\label{sec:arxiv_deploy}
We have deployed the proposed RND-all method on AMiner to disambiguate daily papers from arXiv.org. 
Practically, for each name in the paper to be disambiguated, instead of the adopted name blocking strategy, i.e., moving the last name to the first or preserving all name initials but the last name, we adopt Elastic-Search\footnote{https://www.elastic.co} to perform the online fuzzy search. Finally, we apply RND-all to estimate the similarity between each candidate author and the target paper. To solve NIL cases that there are no right authors, we pre-defined a threshold and return the candidate with the highest score exceeding the threshold as the right author on AMiner.

\subsection{Dataset Statistics.}
\label{subsec:combine}
The detailed data statistics are shown in \figref{tb:overall_sta}.

\begin{table}
	\newcolumntype{?}{!{\vrule width 1pt}}
	\newcolumntype{C}{>{\centering\arraybackslash}p{8em}}
	\caption{
		\label{tb:overall_sta}. Statistics of prevailing manually-labeled name disambiguation datasets. \textmd{Fewer names with more authors and papers indicates the dataset with higher ambiguity.}
	}
	\footnotesize
	\centering 
	\renewcommand\arraystretch{1.0}
	\begin{tabular}{@{~}l?@{~}*{1}{cccC}@{~}}
		\toprule
		\textbf{Datasets} &\textbf{\#Names} &\textbf{\#Authors} &\textbf{\#Papers} &\textbf{Source}
		\\
		\midrule
		Song-PubMed
		&36 &385 &2,875 & \tabincell{c}{PubMed \\ (Biomedicine)}\\
		\midrule
        GS-MEDLINE 
        & - & - & 3,756 & \tabincell{c}{PubMed \\ (Biomedicine)}
        \\
        \midrule
        Han-DBLP
        &14 & 479 & 8,453 & \tabincell{c}{DBLP \\ (Computer Science)}
        \\
        \midrule
        Qian-DBLP
        &680 & 1,201 & 6,783 & \tabincell{c}{DBLP\\ (Several Domains)}
        \\
        \midrule
        Tang-AMiner
        &110 & 1782 & 8386 & \tabincell{c}{AMiner \\ (General Domains)}
        \\
        \midrule
        SCAD-zbMATH
        & 2919 & 2946 & 33,810 & \tabincell{c}{zbMATH \\ (MATH)}
        \\
        \midrule
        Zhang-AMiner 
        & 100 & 12,798 & 70,258 & \tabincell{c}{AMiner \\ (General Domains)}
        \\
        \midrule
        INSPIRE
        & \textbf{12,458} & 36,340 & 360,066 & \tabincell{c}{INSPIRE\\(Physics)}
        \\
		\midrule
		\textbf{\da} 
		& 2,495	& \textbf{72,609}	& \textbf{1,102,249} & \tabincell{c}{AMiner \\ (General Domains)} \\

		\bottomrule
	\end{tabular}
	
\end{table}

\subsection{Interactive Annotation Tool}
\begin{figure*}[t]
	\centering
	\includegraphics[width=\textwidth]{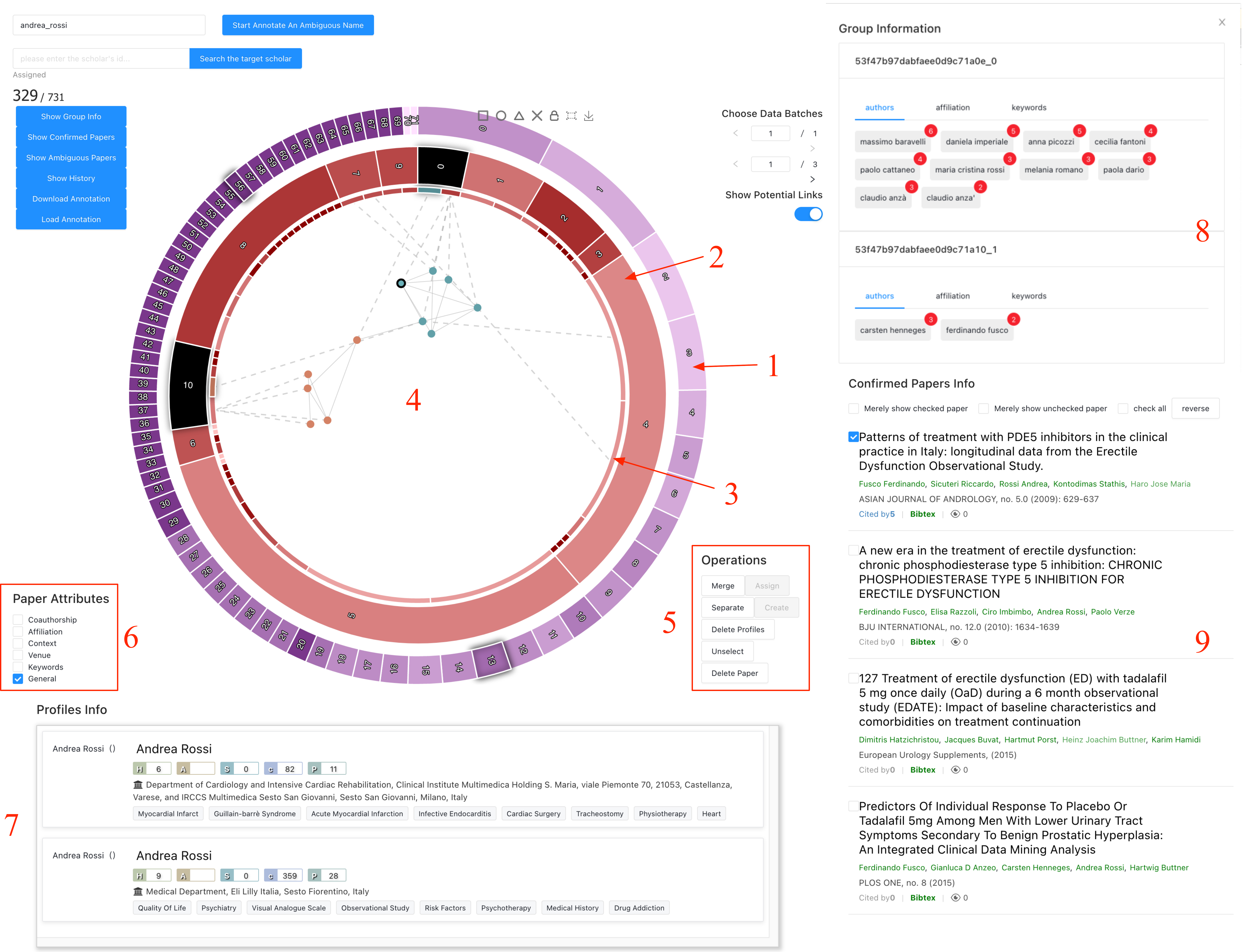}
	\caption{\label{fig:vis_tool} A toy annotation example for annotating authors with author name ``Andrea Rossi''.}
\end{figure*}

\figref{fig:vis_tool} depicts the framework of the designed interactive annotation tool, which consists of two main parts, i.e., the annotation panel and the information panel. 

\ipara{Annotation Panel.} The first 3 regions construct the ring with three stack layers. Specifically, the outer layer, i.e., region ``1'', shows the collected unassigned papers with the target author named ``Andrea Rossi''. Each block in the middle layer, i.e., region ``2'', represents the author named ``Andrea Rossi''. 
To facilitate annotators to grasp the global relationships among papers, we adopt the clustering methods to partition papers within each author into several groups, as shown in the region ``3''. 
Each group contains papers with similar attributes, such as those are published in the same venue, coauthored by the same authors, etc.  By clicking the authors in the region ``2'', we can see the inter-connections, i.e., the dotted line, and intra-connections, i.e., the solid line among papers. Annotators can freely select different attributes via the region ``6''. Then, annotators can perform operations via region ``5''.

\ipara{Information Panel.} Regions ``7'', ``8'', and ``9'' provide comprehensive information about the selected papers and authors, which support annotators to conduct accurate operations. Among these, the region ``7''  presents the profile comparisons among selected authors. Then, region ``8'' shows more detailed comparisons between selected authors, such as coauthors, affiliations, and keywords. Finally, region ``9'' supply the complete information of selected papers.

Overall, the interactive annotation tool not only provides convenient atomic operations to improve the efficiency of annotators, but also prepares comprehensive information to support them make decisions. With the help of the effective visualization tool, we plan to annotate and update more datasets to \sda in the future. 


\end{document}